\documentclass{ieeeaccess}
\usepackage{cite}
\usepackage{amsmath,amssymb,amsfonts}
\usepackage{algorithmic}
\usepackage{graphicx}
\usepackage{textcomp}

\usepackage{array}
\def\BibTeX{{\rm B\kern-.05em{\sc i\kern-.025em b}\kern-.08em
    T\kern-.1667em\lower.7ex\hbox{E}\kern-.125emX}}

\ifCLASSOPTIONcompsoc
\usepackage[caption=false, font=normalsize, labelfont=sf, textfont=sf]{subfig}
\else
\usepackage[caption=false, font=footnotesize]{subfig}

\begin{document}
\history{Date of publication xxxx 00, 0000, date of current version xxxx 00, 0000.}
\doi{10.1109/ACCESS.2017.DOI}

\title{Fast CRDNN: Towards on Site Training of Mobile Construction Machines}
\author{\uppercase{Yusheng Xiang}\authorrefmark{1,2,5}, \IEEEmembership{Student Member, IEEE},
\uppercase{ Tian Tang\authorrefmark{1}, } 
\uppercase{ Tianqing Su}\authorrefmark{3},\IEEEmembership{Member, IEEE},
\uppercase{ Christine Brach}\authorrefmark{2},\IEEEmembership{Member, IEEE},
\uppercase{ Libo Liu}\authorrefmark{4},
\uppercase{ Samuel Mao}\authorrefmark{5},
\uppercase{ Marcus Geimer}\authorrefmark{1},\IEEEmembership{Member, IEEE}
}
\address[1]{Institute of Vehicle System Technology, Karlsruhe Institute of Technology, Karlsruhe, 76131 Germany (e-mail: marcus.geimer@kit.edu)}
\address[2]{Division of Mobile Hydraulics, Robert Bosch GmbH, Elchingen, 
89275  Germany (e-mail: christine.brach@boschrexroth.de)}
\address[3]{Institute of Communication Technology, Technical University of Braunschweig, 38106 Braunschweig, Germany 
(e-mail:  t.su@tubs.de)}
\address[4]{Institute of Energy Conversion and Storage, Ulm University, Ulm,  
89081 Germany (e-mail: libo.liu@uni-ulm.de) }
\address[5]{Department of Mechanical Engineering, University of California at Berkeley, Berkeley, CA, 94720, USA (e-mail: ssmao@berkeley.edu) }


\markboth
{Author \headeretal: Preparation of Papers for IEEE TRANSACTIONS and JOURNALS}
{Author \headeretal: Preparation of Papers for IEEE TRANSACTIONS and JOURNALS}

\corresp{Corresponding author: Yusheng Xiang (e-mail: yusheng.xiang@partner.kit.edu).}

\begin{abstract}
The CRDNN is a combined neural network that can increase the holistic efficiency of torque based mobile working machines by about 9\% by means of accurately detecting the truck loading cycles. On the one hand, it is a robust but offline learning algorithm so that it is more accurate and much quicker than the previous methods. However, on the other hand, its accuracy can not always be guaranteed because of the diversity of the mobile machines industry and the nature of the offline method. To address the problem, we utilize the transfer learning algorithm and the Internet of Things (IoT) technology. Concretely, the CRDNN is first trained by computer and then saved in the on-board ECU. In case that the pre-trained CRDNN is not suitable for the new machine, the operator can label some new data by our App connected to the on-board ECU of that machine through Bluetooth. With the newly labeled data, we can directly further train the pre-trained CRDNN on the ECU without overloading since transfer learning requires less computation effort than training the networks from scratch. In our paper, we prove this idea and show that CRDNN is always competent, with the help of transfer learning and IoT technology by field experiment, even the new machine may have a different distribution. Also, we compared the performance of other SOTA multivariate time series algorithms on predicting the working state of the mobile machines, which denotes that the CRDNNs are still the most suitable solution. As a by-product, we build up a human-machine communication system to label the dataset, which can be operated by engineers without knowledge about Artificial Intelligence (AI).
\end{abstract}

\begin{keywords}
Transfer learning, Deep learning, Construction machine, Human machine communication, Multivariate time series classification, Internet of Things

\end{keywords}

\titlepgskip=-15pt

\maketitle

\section{Introduction}
\label{sec:introduction}
\PARstart{I}{n} the previous study \cite{Xiang.2020d}, the CRDNN shows excellent performance in detecting the Y cycles of primary-torque-based mobile machines. To date, we believe that CRDNN is a promising method to solve the problem. Firstly, it is an offline approach that can be an order of magnitude faster than the other online learning methods. Also, it achieves a better performance on the challenging dataset by taking the time-series signal sequence into account. However, due to the diversity of mobile machines and driver behaviors, the accuracy of prediction is not always so satisfying even the CRDNN is used. The performance of CRDNN decreases when it faces measured data from a driver with totally unseen behaviors, which means the distribution of data gathered from the new machines and drivers are different from the previous dataset used to train the CRDNN. The reasons are apparent. First and foremost, the CRDNN is an offline learning method that can not automatically adapt to the new tasks after it has been trained. Also, the gather of all the data in every scenario for the initial training is still challenging and, of course, economically impossible. Therefore, in this paper, we utilize the transfer learning and IoT technology to solve the problem. The pre-trained CRDNN will be further trained in case that the machines or drivers have totally different features, and the recognition system can then reach the expected performance. Apparently, establishing the communication interface between humans and machines plays a vital role in this approach. Therefore, this communication interface is also introduced in this paper.

\Figure[ht!](topskip=0pt, botskip=0pt, midskip=0pt)[width=3.3in]{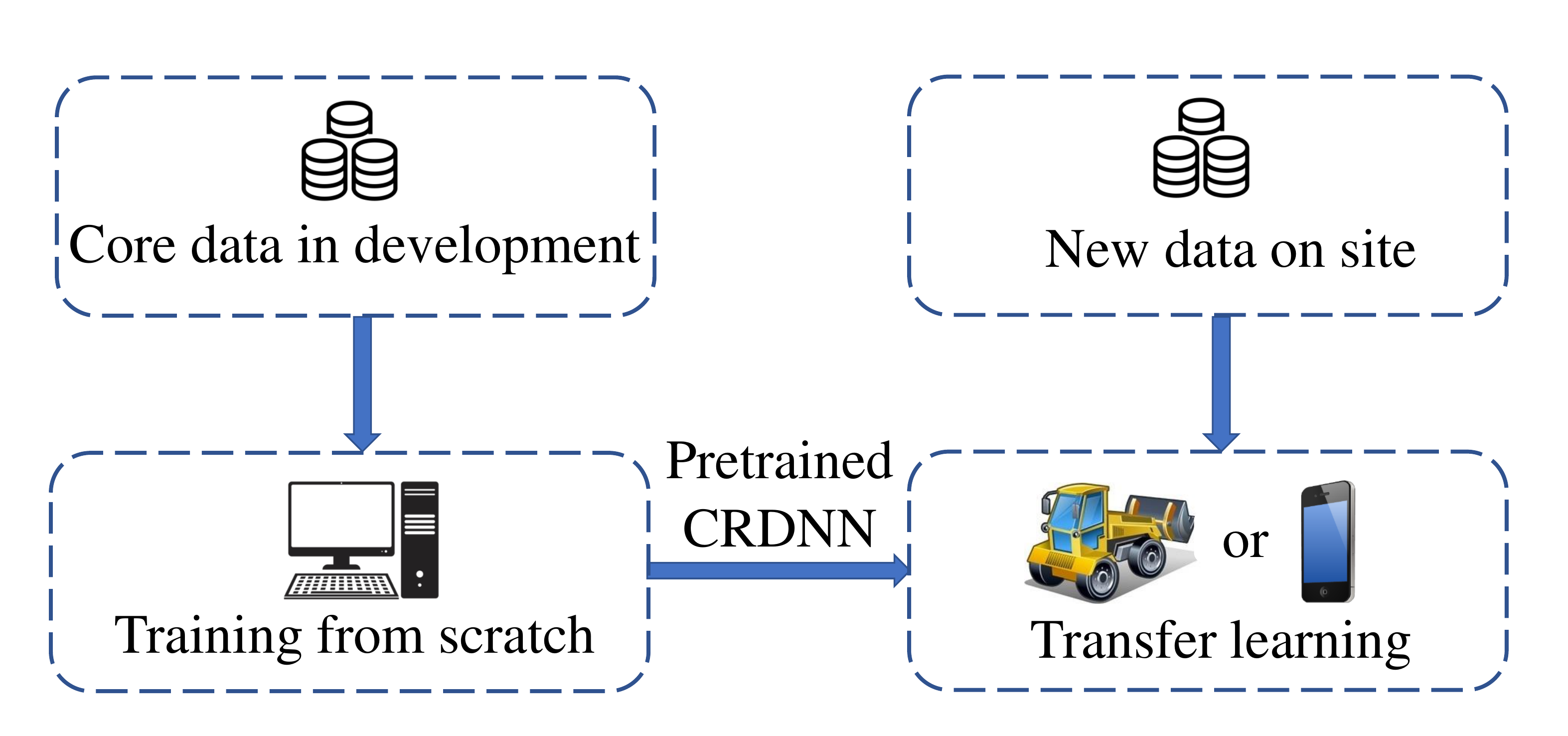}
{Here the core data is a large dataset that contains 119 Y cycles data from many wheel loaders. This core dataset is used to train the base network. Thanks to this base network, we can then use transfer learning to adapt the weights in this base network with the new data to improve the generalization ability, easy and quick. The method is proposed to solve the problem pointed out by many machine learning researchers, the distribution of the source data may differ from the target data since the collection of a comprehensive dataset is, in many cases, impossible.\label{fig:problem}}

The main contributions of this paper can be sum up as the following points:
\begin{itemize}
    \item We compare the performance of the selected CRDNNs to another commonly used SOTA solution in the field of Time Series Classification (TSC), and show that CRDNNs is more suitable for the Y cycles detecting task
    \item We proposed that transfer learning should be used to enhance the generalization capability of CRDNNs
    \item We recommend CRDNN with 2 LSTMs as the base network based on its micro F1,  back- and forward propagation duration so that the networks can be further trained directly on the working site using transfer learning
    \item We design an easy human-machine communication system for the data exchange between human and mobile machines
    \item We proposed an approach to label the slip windows which can reduce the delay between the state occur and the state can be correctly predicted
\end{itemize}

The rest of this paper is organized as follows. Section II briefly introduces the prerequisite and background knowledge in fields of mobile machines, IoT, and time series classification to understand this paper since our readers might come from these three fields. Next, the existing problems and proposed solutions are illustrated in Section III. Then, the reasons why we adopt these solutions are provided in Section IV. After that, in Section V, we describe the connection system between the human and the mobile machines. In section VI, we show how the measurement setup. Followed by section VII, we compare the variations of CRDNNs with the SOTA TSC solution, and the performance of different transfer learning methods.  Finally, Section VIII gives conclusions and envisions the outlook.

\section{Background knowledge}

\subsection{The future mobile construction machines drive train system }

Currently, the mobile machines use the flow-based controlled drive train, which controls the vehicle speed by the volume flow pass the hydraulic motor and thus the vehicle velocity \cite{Bauer.2019}. The advantages of such a drivetrain solution are due to the decoupling of the engine and the vehicle speed \cite{Xiang.2020b}. However, the efficiency of this concept can even lower than 10\% \cite{Ge.2018} in many applications. Thus, many variations based on these concepts have been drawn \cite{Ge.2017}. Based on our literature analysis, we find the research focus of the scientists in the field of mobile construction machines goes to the torque based controlled concept \cite{Vukovic.2016,Xiang.2020c,Sprengel.2015}. The initial proposal to introduce the torque control concept consists of higher holistic efficiency, flexible system architecture due to modulation, and more suitable for the employment of a hybrid system. Apparently, different control concept leads to different system layout and corresponding internal sensors selection. Since torque-controlled mobile machines may win the competition in the long term,  we focus on the technologies that can be used on the torque-based mobile machines in this paper, especially the primary torque-based control introduced by Bosch Rexroth AG in 2018 \cite{Mutschler.2018} for hydrostatic mobile machines. Since the measured variables in the primary torque concept can also be interpreted as secondary control concept, our algorithm can be principally adapted to the secondary controlled mobile machines with some further works.

\subsection{Y cycles detection of wheel loaders}


\Figure[ht!](topskip=0pt, botskip=0pt, midskip=0pt)[width=7in]{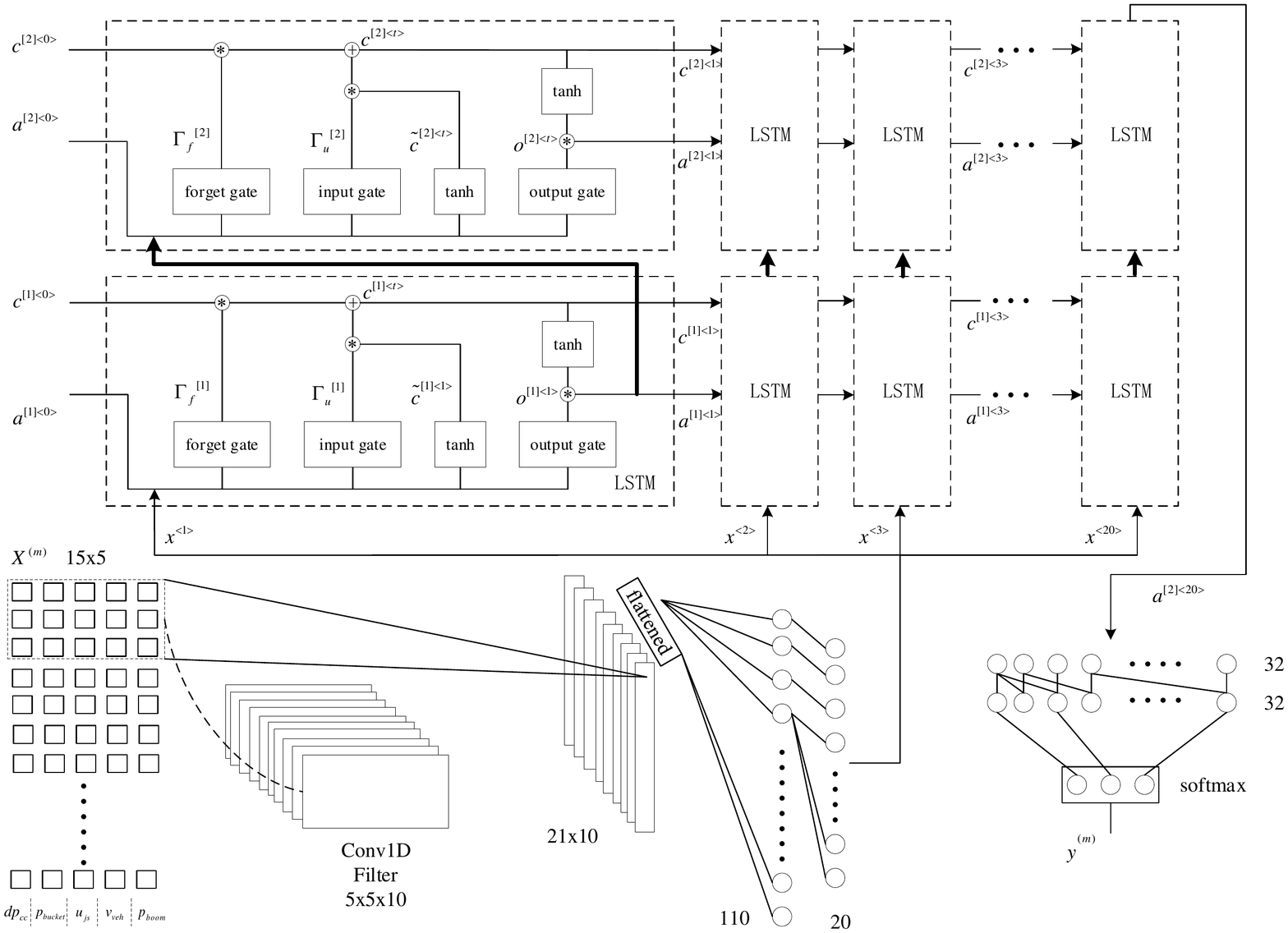}
{The detailed description of CRDNN with two layers LSTM. The notation we use is based on the Stanford University deep learning Lecture Notes, where $ c^{[i]<j>}$ denotes the cell in $i_{th}$ layer and for $j_{th}$ time series.\label{fig: CRDNN}}

The Y cycles are the most typical working process of wheel loaders. The performance during the Y cycles has a decisive effect on the holistic performance of the mobile machine. In our previous works, the CRDNN is validated as an excellent AI tool to solve the detection problem so that the corresponding operation strategy can be easily designed, and then the machines can recuperate the energy to improve the holistic performance. A concrete description of how we regenerate the energy  can be found in \cite{Xiang.2020d} with more details. Also, the working process detection is used as a vital criterion to predict the intention of the drivers.

\subsection{What is CRDNN?}

As aforementioned, CRDNN is a combined neural network that combines the Convolutional Neural Networks (CNN), Recurrent Neural Networks (RNN), and Dense Neural Networks (DNN). The combination brings the advantages of different kinds of neural networks together\cite{Sainath.2015}.  Pressure inside the bucket ($p_{bu}$), vehicle velocity ($v_{veh}$), vehicle direction signal on the joystick ($u_{js}$), pressure inside of closed-circuit drivetrain ($p_{cc}$), and Pressure inside of bucket ($p_{bo}$) are collected during the wheel loaders are working in Y cycles. We labeled the data with corresponding working state, traveling ($e_0$), loading ($e_1$), and unloading ($e_2$). We then trained our neural network on the computer with these ground truth data. In order to find out the best model for the task, we have explored many different kinds of networks, such as CNNs, RNNs, DNNs, and their combinations. Among these neural networks, the combined neural networks CRDNN with two LSTM layers performs excellent test accuracy with relatively low training parameters. Moreover, the robustness of this model to the small amount of mislabelled data is also the reason for the final selection. We saved the trained CRDNN in the on-board ECU, and CRDNN can rapidly identify the working state with high precision and recall. The model is built with Kereas API in Tensorflow \cite{Abadi.2016}. A more detailed description of how we built up the dataset and the CRDNN can be found in our previous study \cite{Xiang.2020d}.

\subsection{Long Short Term Memory Fully Convolutional Network: a SOTA solution for TSC tasks}

Long Short Term Memory Fully Convolutional Network (LSTM-FCN) is designed for classifying univariate time series \cite{Karim.2018}. In order to apply this network to the multivariate time series classification problem, Karim extended the Squeeze-And-Excite (SAE) block to the case of 1D sequence models and augmented the fully convolutional blocks of the LSTM-FCN model to improve classification accuracy \cite{Karim.2019}. The network architecture is shown in Fig. \ref{fig4.12}.

\Figure[ht!](topskip=0pt, botskip=0pt, midskip=0pt)[width=3.3in]{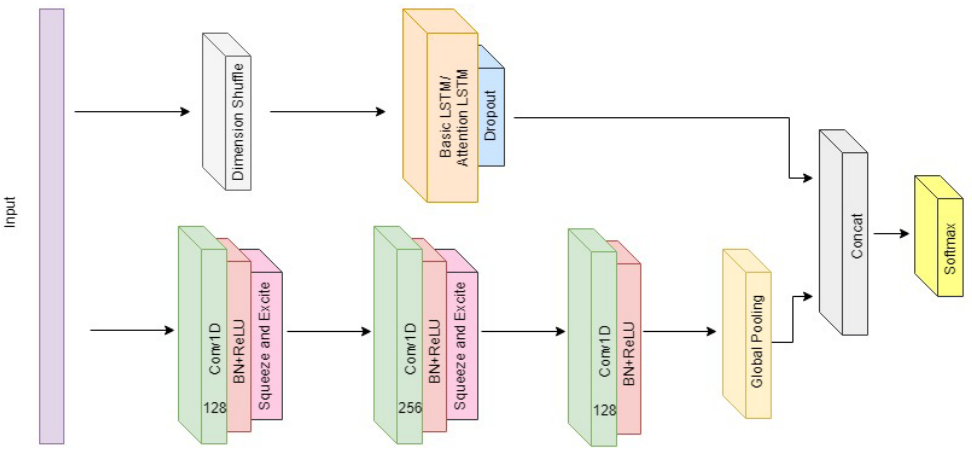}
{LSTM-FCN with squeeze-and-excite block \cite{Karim.2019} \label{fig4.12} }

Fully Convolutional Networks (FCN) have proven to be an effective learning model for time series classification problems \cite{Wang.20175142017519}, which comprised of three temporal convolutions, are typically used as feature extractors. Global average pooling \cite{Lin.2013} is used to reduce the number of parameters in the model before classification. The SAE is added after the FCN block which adaptively recalibrates the input feature maps \cite{Karim.2019}. 

This architecture has been tested on 35 benchmark datasets for TSC, and it outperforms the other SOTA models on at least 28 data sets. Thus, we would like to compare our CRDNN with this algorithm for the task of detecting mobile machines' Y cycles.

\subsection{Wireless connection of mobile construction machines}

To achieve the smart working site, effective communication among mobile machines is an inevitable vital step. Since the mobile machines are very likely to work at a place where there is outside of the coverage of the base station, we utilized the ad-hoc network as the first version for the fleet management of the mobile machines \cite{Xiang.2020}. In that paper, although the realtime communication system is proposed, the bidirectional communication between human and mobile construction machines is still a gap. Recently, many other scientists also emphasize the value of setting up the management system between operators and machines \cite{VillaHenriksen.2020}. However, they did not consider the rapid development of the new technology on mobile smartphones and consequently did not develop core functions on the smartphone. Based on the research from Ignatov, the capability of the system on a chip (SoC) on cell phone grows extremely fast and research almost 40\% velocity of Geforce GTX 1060 in terms of processing images \cite{Ignatov.2018}. Hence, we would like to build up a connection between cell phones and our mobile construction machines to take advantage of the cell phone SoC industry's development. The top SoCs until April 2020, A13 from Apple Inc, Snapdragon 865 5G from Qualcomm, Kirin 990 5G from Huawei, Exynos 990 5G from Samsung, claim that their SoCs can be about 20\% faster compared to their last generation published in the last year. Also, the newest version SoCs equip with GPU to enhance the capability to deal with artificial intelligence tasks. All of them have Bluetooth 5.0 modules that can easily connect to the mobile construction machines onboard ECU. Apparently, the development of the computational performance of SoCs is much faster than the onboard ECU.

\section{Problem statement and brief description of the solution}

As our first version of CRDNNs, the CRDNNs can easily achieve predictive accuracy of about 98\% based on the dataset of 119 Y cycles, which reaches human-level performance. However, when we consciously change the equipment, especially the shovel, of those mobile machines and test the CRDNNs, the performance is degraded to an unacceptable level. Fig. \ref{fig: problem} illustrates the performance of CRDNN when it faces measurement data from a driver with totally unseen behaviors, and the implements has been changed. The reason for that is the training data and the test data have a different distribution in both marginal distribution and conditional distribution.

\Figure[ht!](topskip=0pt, botskip=0pt, midskip=0pt)[width=3.3in]{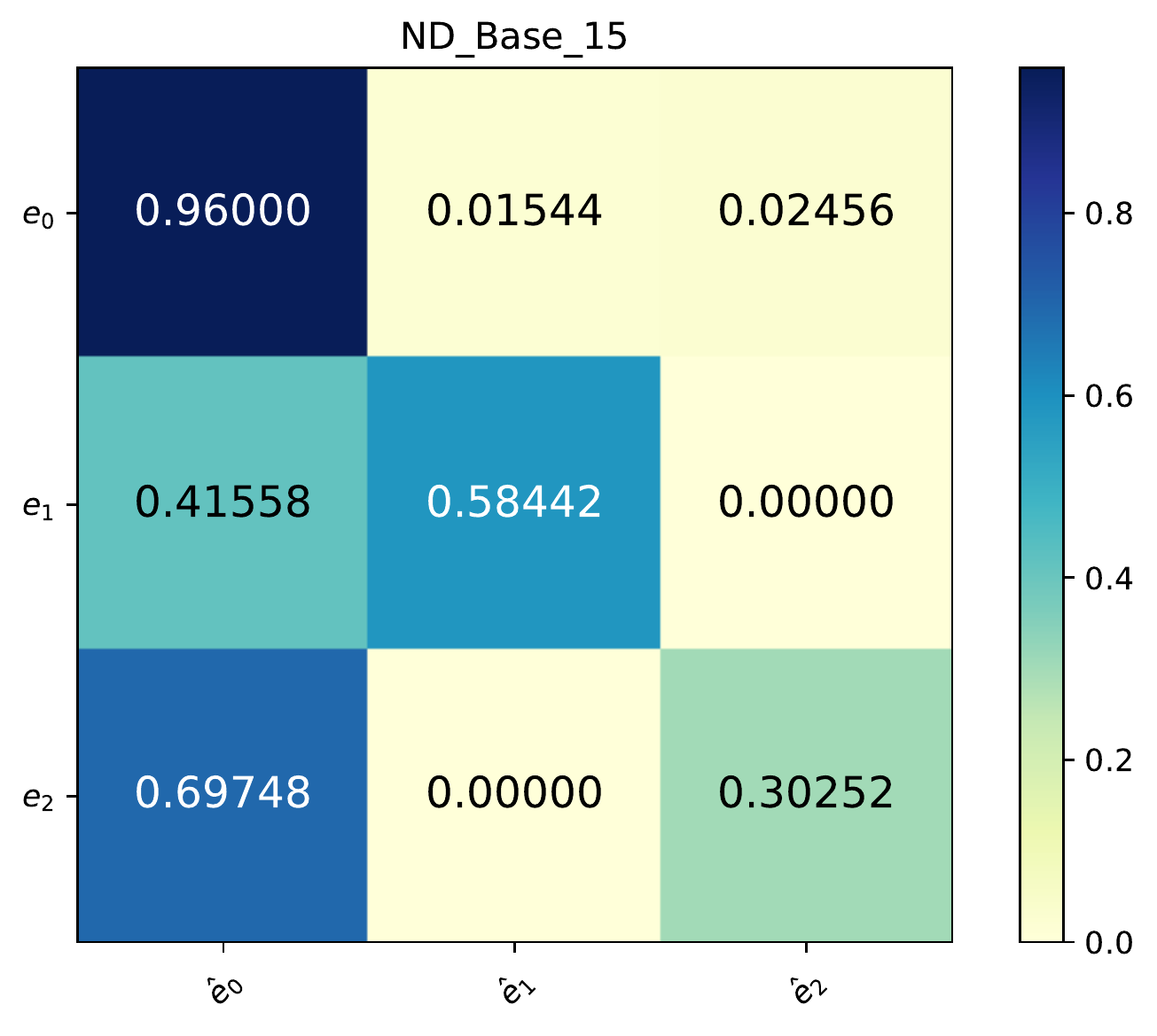}
{The confusion matrix of CRDNN on new data. The $e_i$ are the ground truth and the $\hat{e}_i$ is the predictive state. As in our previous work defined, the $e_{1,2,3}$ denote the state travelling, loading, and unloading, separately.  \label{fig: problem}}

In addition, since the mobile machines are rent for construction tasks and count money by time, the robustness of the machines and the algorithms on the machines is a matter that cannot be negotiated for the contractors. Thus, either an approach that can always guarantee the performance of the algorithm without adjustment or an approach that only requires rapid and easy calibration is needed as a complementary solution. Even worse, OEMs are reluctant to share their data with each other resulting in a lack of training data for all of them. Based on the facts and challenges we analyzed, we select the approach of offline learning with online adaption. Concretely, instead of sharing the real measured data, transfer learning allows them to further train the pretrained base neural networks with a small new set and thus have a similar effect as they gain a series of data and train the neural networks.

As we know, data plays a critical role in deep learning. A large and highly diverse dataset improves the capability of machine learning methods. Also, the same distribution and feature between the training data and test data are a guarantee for the excellent performance when the neural networks are applied in practice. However, in the real world, there are many different kinds of construction machines and workplaces, which may lead to the change of the data distribution. Since the collection of the dataset from all kinds of construction machines is almost impossible, we adopt the transfer learning method to guarantee the same data distribution of the training and test data.  Since there must be some similarities between the data we collected from the previous wheel loaders and the new machines, we can do fine-tuning with labeling a few datasets on the working site, and it will only take a few training steps to achieve the satisfying prediction results. Thus, it is not computationally expensive and can be trained directly by the onboard ECU or smartphone. Notice that, whether the new data should be trained on the onboard ECU or the SoC in the cell phone is depending on the capability of them and the bandwidth of the connection. At present, we recommend further train the CRDNN on the onboard ECU since the transmit of the data from mobile construction machines to cell phones has a more massive amount of data as in reverse. However, the approach introduced in this paper can be easily adapted to the version that trains the CRDNN on the cell phone at the time when the data transmission is proved as no more a problem.

\section{Why transfer-learning based supervised learning?}

Traditional machine learning performs well by using training data and testing data with the same input feature space and the same data distribution. When there is a difference in data distribution between the training data and test data, the results of a predictive learner is likely to be degraded \cite{Shimodaira.2000, QuioneroCandela.2009,Zhou.2014}. In certain scenarios, obtaining training data that matched the feature space and predicted data distribution characteristics of the test data could be difficult and expensive. Therefore, there is a need to create a high-performance learner for a target domain trained from a related source domain. This is the motivation for transfer learning \cite{Zhuang.2019}. Transfer learning is used to improve a learner from one domain by transferring information from a related domain \cite{Bishop.2006, Dai.2007} .

Since the transfer learning is a rapid developing subject, the terminology and definition have currently no consistency. In this paper, we use the mathematical definition from Pan for further discuss, who defined that $\mathcal D_s= (X_s, P(X_s))$ as source domain, $\mathcal D_t= (X_t, P(X_t))$ as target domain, $\mathcal  T_s=(X_s, f_s(\cdot))$ as source task, and  $\mathcal T_t=(X_t, f_t(\cdot))$ as target task. Transfer learning aims to enhance the learning of the target predictive function $f_t(\cdot)$ in $\mathcal D_T$ using the knowledge in $\mathcal D_s$ and $\mathcal T_s$, where $\mathcal D_s \neq \mathcal D_t$, or $\mathcal T_s \neq \mathcal T_t $ \cite{Pan.2010}.


In the past decade, transfer learning has been successfully implemented in the fields of image recognition \cite{Csurka.2017,Rozantsev.2019} and Natural Language Process \cite{Jiang.2007}. In contract, scientists in the field of TSC believe that there has a lot of things should be proven or improved \cite{Gamboa.2017}.  It is only recently that deep learning was proven to work well for some TSCs \cite{LeGuennec.2016}. However, unlike image recognition \cite{Russakovsky.2015}, the lack of a sizeable general-purpose dataset in TSC limits the development of transfer learning in TSC. Another well-known problem by implementing the transfer learning on the TSC task is the negative transfer. As we know, if one is good at handball, she or he can learn how to play basketball faster than the others who never played handball before. The reason is apparent: the knowledge about how to play handball and basketball well are similar. However, people usually have a negative evaluation of the people who give them a bad first impression ($\mathcal D_s$), no matter how other people change ($\mathcal D_t$). For the latter example, the first knowledge ($\mathcal D_s$) does not contribute to the correct prediction ($f_t(\cdot)$) and indeed has an adverse effect. This is a negative transfer. The negative transfer and how transferable are features are still very active research domain \cite{Yosinski.2014}. Fawaz has revealed that transfer learning can both improve or degrade the model prediction depending on the source dataset ($\mathcal D_s$) \cite{Fawaz.}, by testing the performance of Fully Convolutional Network (FCN)
algorithm \cite{Wang.20175142017519} with transfer learning and from scratch on a series of dataset. As the best of the author's know, the consensus is that transferring models between similar datasets improves the $f_t(\cdot)$ performance.  In contrast, Rosenstein empirically showed that if two tasks are too dissimilar, then brute-force transfer may hurt the performance of the target task \cite{Rosenstein.2005}. Thus, Mahmud proved some theoretical bounds by analyzing the case of transfer learning using Kolmogorov complexity \cite{Mahmud.2008}. Furthermore, some previous works have been exploited to analyze relatedness among tasks by using clustering techniques, which provide the guideline about how to automatically avoid negative transfer \cite{BenDavid.2003, Bakker.2003}. Keogh shows that dynamic time warping is a robust distance measure for time series, which can thus evaluate the similarity of the dataset \cite{Keogh.2005}. Based on the literature recherche in the field of transfer learning, we can conclude that the more similarities between the ($\mathcal D_s$) and ($\mathcal D_t$), the better transfer learning can perform.

There are different strategies and implementations for solving a transfer learning problem. The majority of the homogeneous transfer learning solutions employ one of three general strategies which include trying to correct for the marginal distribution difference in the source $P(X_s)  \neq P(X_t)$, trying to correct for the conditional distribution difference in the source $ P(Y_s|X_s)  \neq   P(Y_t|X_t )$, or trying to correct both of them \cite{Weiss.2016}. 

Some similar use cases for TSC with transfer learning can be found in many previous studies. For example, Hu proposed first to train a model on the historical wind-speed data of an old farm and fine-tune it using the data of a new farm \cite{Hu.2016}. In addition, Peng propose a transfer-learning based approach to establish an anomaly detection model for dangerous actions of aircraft testing flights \cite{Xiong.20189212018923}. A transfer learning-based bi-directional long short-term memory model is proposed to predict the air quality by Ma \cite{Ma.2019}. The success of the implementation of transfer learning on TSC tasks encourages us to follow this concept. 

In our transfer learning task, the data we used to pre-train the base network from scratch is the source domain ($\mathcal D_s$), while the data we collect from the new machines are the target domain ($\mathcal D_t$). Apparently, the solution to this problem is to correct both the marginal distribution and the conditional distribution difference in the source. It can be referred to as a parameter-transfer approach, which assumes that the source tasks and the target tasks share some parameters or prior distributions of the hyper-parameters of the models. Our transfer learning approach is to recompute the trainable parameters in the neural network. The architecture of the base network will be kept the same.

\Figure[ht!](topskip=0pt, botskip=0pt, midskip=0pt)[width=6.7in]{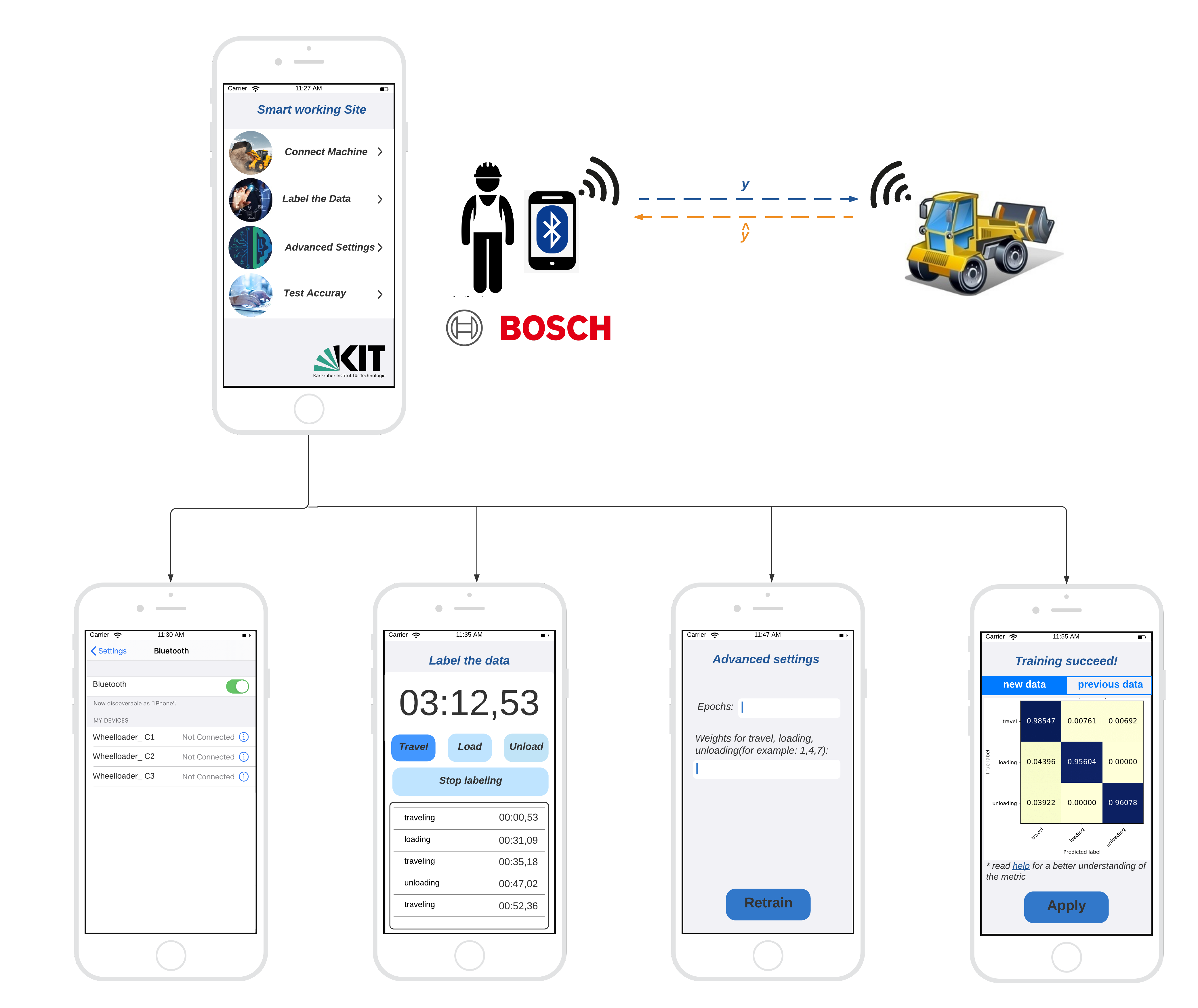}
{The sketch of the human-machine communication App system \label{fig:app}}

Another potential approach is also mentioned, which could be used to detect Y cycles: semi-supervised sequence learning, which leverages the unlabeled data to further improve the predictive accuracy \cite{Berthelot.2019}. However, the performance of semi-supervised learning is quite difficult to outperform supervised learning \cite{vanEngelen.2020}. This method is usually adopted for the private data task, where label the data is prohibited \cite{Dai.2015}. In the case of detecting Y cycles, obtain the new data is actually only a technical problem, and the data must be much easier to get in the era of IoT; thus, we would use supervised learning instead. To achieve the transfer-learning based supervised learning, we have designed a connection system between the mobile machines and human using smartphone.

\section{Connection system design}

\subsection{Choice of Wireless Communication Technology}\label{AA}
There are mainly four common short-range wireless communication technologies in the field of IoT, namely Near-field communication (NFC), Radio-frequency identification (RFID), Bluetooth, WIFI. The comparison of their main specifications are shown in Tab.  \ref{tab_6.1}. 
\begin{table*}[ht]
\caption{the main specifications of the wireless communication technology}
\begin{center}
\begin{tabular}{m{3cm}<{\centering}|m{2cm}<{\centering}m{2cm}<{\centering}m{3cm}<{\centering}m{2cm}<{\centering}}
\hline\hline
\textbf{Specifications} & \textbf{NFC}&\textbf{RFID}&\textbf{Bluetooth5.0}&\textbf{WIFI}\\\hline
\textbf{Maximum Coverage Range} & \textbf{10 cm}&\textbf{3 m}&\textbf{100 m}&\textbf{100 m}\\

\textbf{Radio frequency} & \textbf{13.56MHz}&\textbf{varies}&\textbf{2.4GHz}&\textbf{2.4GHz, 5GHz}\\

\textbf{Communication mode} & \textbf{2-way}&\textbf{1-way}&\textbf{2-way}&\textbf{2-way}\\

\textbf{Data Rate} & \textbf{106,212, 424Kbps}&\textbf{varies}&\textbf{2Mbps}&\textbf{144Mbps}\\

\textbf{Applications} & \textbf{Credit card related payments, E-ticket booking}&\textbf{EZ-Pass, Tracking items}&\textbf{Communication between phone and peripherals}&\textbf{Wireless internet}\\
\hline\hline
\end{tabular}
\label{tab_6.1}
\end{center}
\end{table*}

In order to enhance the generalization capability of CRDNN, we need to get the new labeled data to train the pre-trained base network further. The new data is labeled through the mobile app, which connects ECU through the Bluetooth. With the new labeled data, the network is retrained on the ECU, and the accuracy of the retrained network can be shown in the app. When the test accuracy reached the expectation, the machine can be put into use.

Considering that most machine operators are not specialists in deep learning, we design the interface as naturally as possible. We find that only two tasks must be done manually: labeling the data and check the confusion matrix. The other steps will be done automatically either by the APP or the ECU.

Each of those technologies has its pros and cons, and can be implemented into different scenarios. NFC can be easily used for transactions, but not for on-site training due to the limited range, which is approximately 10 cm. RFID technology provides a reliable, efficient way to transmits the identity of an object \cite{Madakam.2015}, so that it is widely used in the area of the E-ZPass system \cite{Nath.2006}. However, RFID only supports the one-way transmission, and therefore it is not a solution for our use case. Compared to WIFI, Bluetooth has a lower energy consumption and more straightforward hardware implementation \cite{Zhao.2014}. Therefore, we select Bluetooth for our on-site training. 
To date, the latest version in Bluetooth is Bluetooth 5, which is introduced by the Bluetooth Special Interest Group (SIG). This version offers significant enhancements compared to the previous specifications, regarding a broader range up to 200 m, a faster speed up to 2 Mbps, and more robust to interference \cite{Mario.2018}.

\subsection{User Interface of the system}\label{BB}

The on-site training system is presented in Fig. \ref{fig:app}. Following, we are going to describe the process that fine-tuning the model on the onboard ECU.  The system consists of a mobile smartphone for labeling date manually and the mobile construction machine, which is equipped with Bluetooth Low Energy (BLE) transceiver chip for communicating with the mobile device. The construction machine operator installs our ``Smart working Site" app, which is demonstrated in \ref{fig:app}. The app provides four perspectives, namely ``Connect machines", ``Label the Data", ``Advanced Settings", and ``Test Accuracy". At the beginning of the on-site training, the machine operator shall activate the Bluetooth of the smartphone and pair the construction machine, as long as the construction machine is situated within the Bluetooth coverage of the smartphone. In the next, the operator observes and records the construction machine's actions, as the driver starts the construction work. The machines' working states are transmitted to the on-board ECU intermediately, once one action is labeled.  This time series of labels indicates the current action of the machine and is served as ground truth for transfer training of the network. For those who are familiar with neural networks, they can tune the hyperparameters as well as different learning algorithms to retrain the network in the tab of ``Advanced Settings". However, use the model we recommend in this paper can fix most of the problems; thus, we are not suggested to use the advanced function on the smartphone unless the operators are extremely confident. The hyperparameter ``epochs" indicates the number of loops, in which all the training data are fed to the network. The other indicator ``weights" means the priority of each working state to be correctly predicted. As the last step, Onboard ECU retrains the network work and transmits the accuracy back to the app, which is visible in ``test accuracy". Once the performance is satisfying, the retrained neural network is applied to the machine.

\section{Measurement setup}

To simulate the situations which the OEMs are likely to meet, we consciously change the control algorithm of the implement, and also the size of the shovel. In fact, in order to adapt to different tasks, OEMs will modify a different control program to facilitate the driver's operation. Also, the machines have different sizes for the different working sites; among these differences, the most considerable distinction is the shovel sizes. Therefore, our measurement is set up based on these facts. 

Fig. \ref{fig:radlader} shows the mobile machine which we use to gather the new data. Thanks to the dSpace, we can change the control algorithm on this prototype mobile machine with ease. 


\Figure[ht!](topskip=0pt, botskip=0pt, midskip=0pt)[width=3.3in]{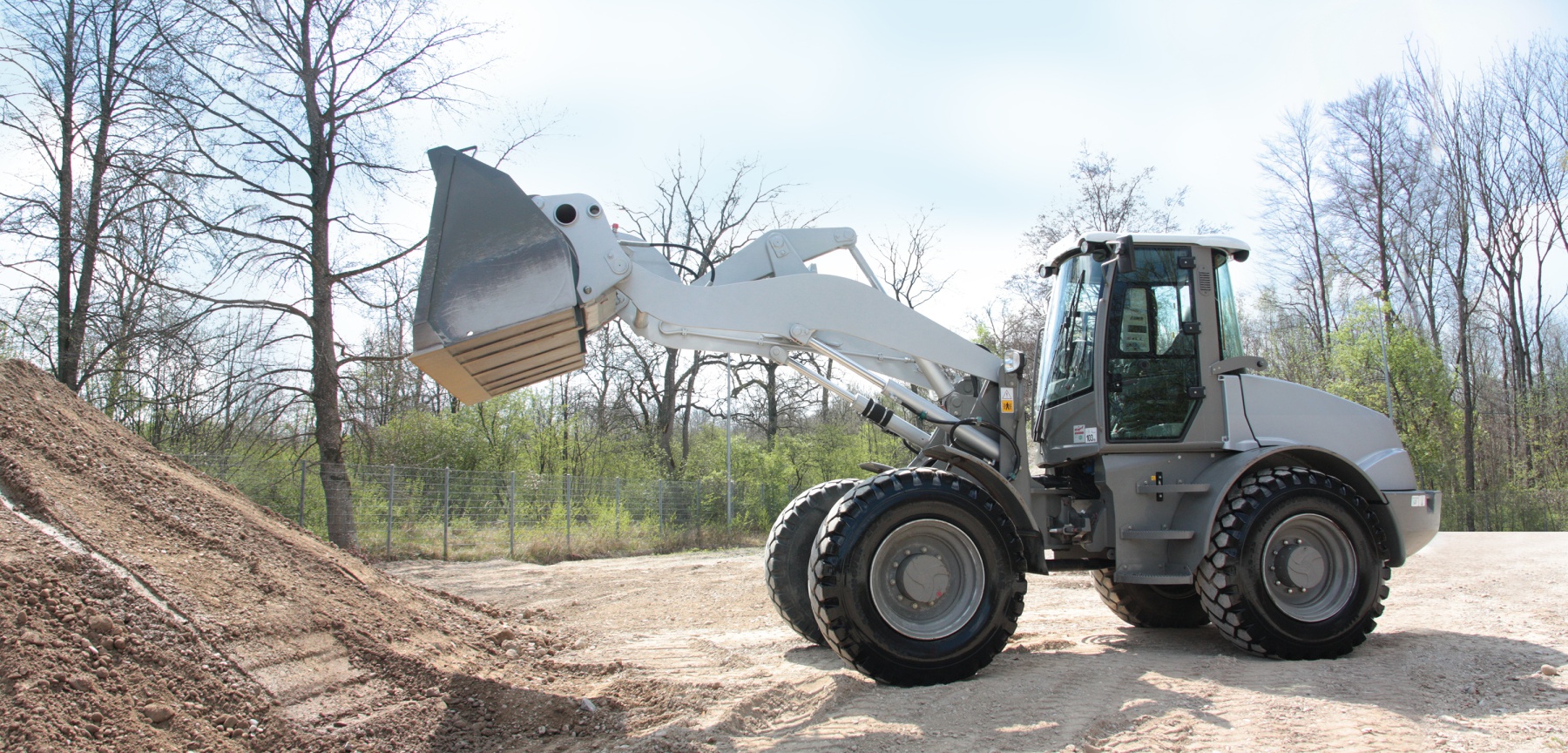}
{The mobile machine used for the measurement data \label{fig:radlader} }


\Figure[ht!](topskip=0pt, botskip=0pt, midskip=0pt)[width=6in]{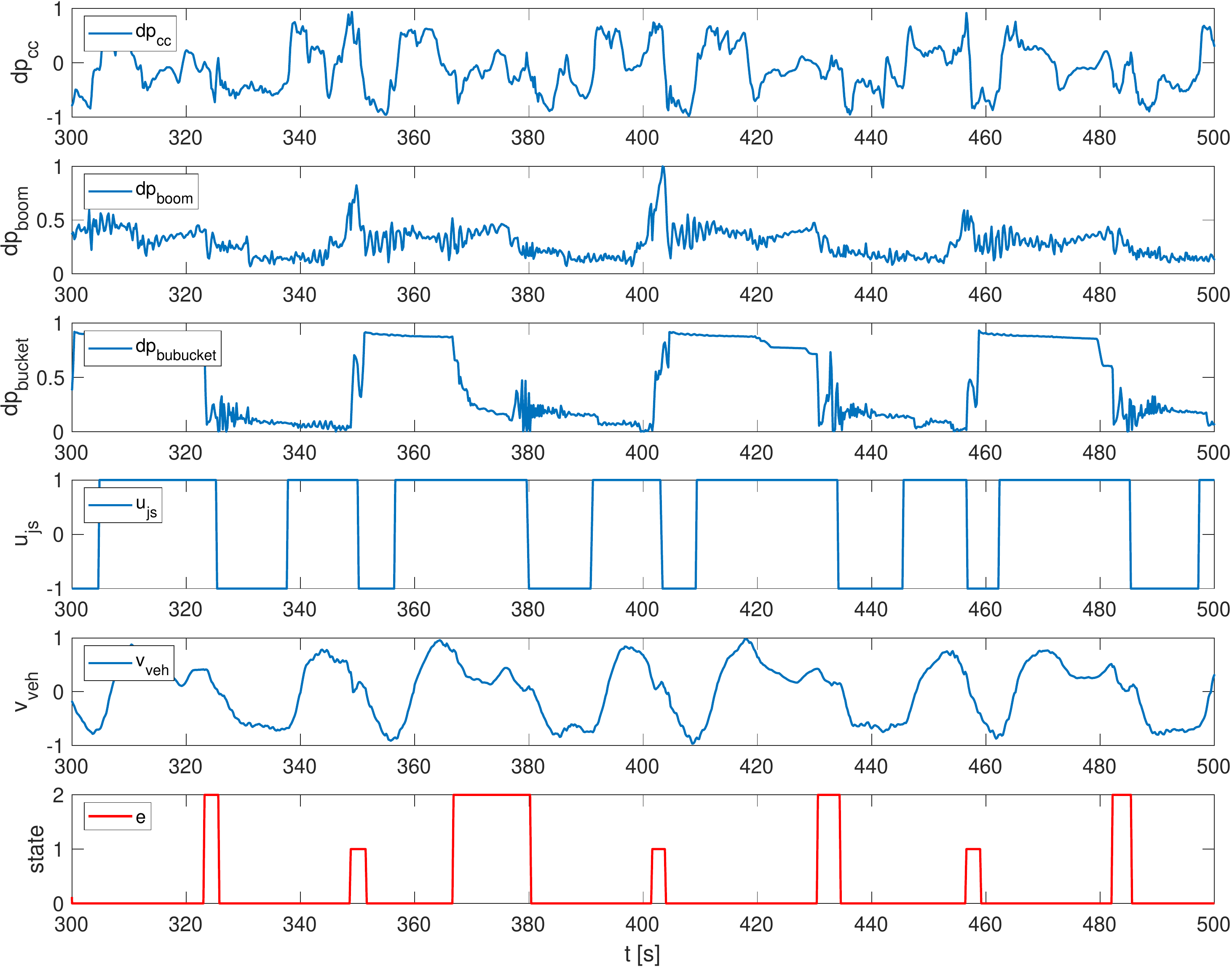}
{New measurement data with a different implement control algorithm and dimension of the shovel. The last subfigure shows the ground truth state of the measured data \label{fig: measurement} }

The newly gathered measurement data, including 24 Y cycles, are partly shown in Fig. \ref{fig: measurement}. By observing the newly gathered data  $\mathcal D_t$, we find that the driver operated joystick differently, compared to the driver who created the original dataset $\mathcal D_s$. The dataset is normalized to accelerate the training process so that the influence of the varying of the shovel dimension might not be shown clearly. 

In order to simulate the fact that different engineers may have divergence on how to label the data since they have different standards or rules, we consciously label the newly gathered data in another way as the previous study. For the new dataset, we label the sample into the state traveling whenever the $dp_{bu}$ is still fluctuating, which is different from the previous approach. Consequently, the distribution of the new dataset has also changed, so the marginal distribution of the source data and target data is much different. 
To sum up, for the new measurement, we purposefully chose a different driver, a different control algorithm for the implement, a shovel in a different size, and a different engineer to label the dataset. Although this makes the task more challenging, we believe it is more approaching to the reality and should be taken into consideration.

\subsection{The sliding windows labeling method}

After the raw data are gathered and labeled, we need to split the time series data into some small sliding windows to train the neural networks. We sample the data in $5Hz$ to avoid overloading the ECU. Obviously, the window sizes affect the system performance; the more significant the window sizes are, the more information will be taken into consideration, and thus more accurate can be expected. However, a larger window size may result in a delay between the state occurs and the machine detects the state. Following, we will illustrate the mechanism of this delay. 

\subsubsection{Labeling the slide windows based on the whole data}

We did not use the state of the last sample data in the sliding window as the state of the slide windows, because the time point where the state changes are vague. Thus, we believe that we should not label the slide windows only based on one sample data in it. Another drawback of only using one sample data is, the consequently labeled sliding windows can make neural network confusion since most of the sample data in this sliding window might indicate another state.  

Here we set the slide windows length as 15, which means a sliding window contains 15 sample data with the label. In order to label these sliding windows, we can calculate the distribution of the samples. In this fashion, the slide windows must have an odd length. In case that one state has the majority, we can then set these windows as this state. For example, if 7th sample data have labeled as loading and 8th sample data are labeled as traveling, the sliding window will be labeled as loading since traveling is the majority.  However, in this case, the state traveling occurs at the 8th sample data, and the machine detects the sliding window as traveling when the 15th sample data is measured. Therefore, a delay exists principally by this method.  The method can be explained by Fig. \ref{fig1} concretely. 


\Figure[ht!](topskip=0pt, botskip=0pt, midskip=0pt)[width=3.3in]{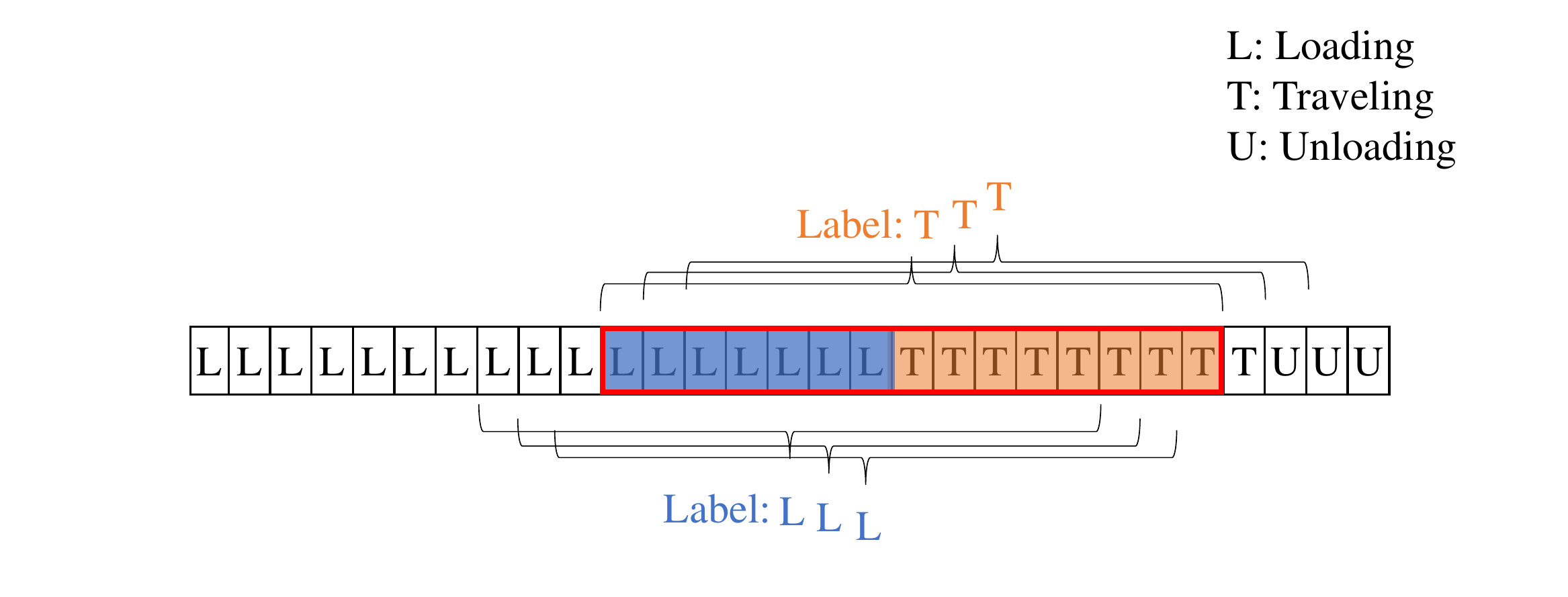}
{The diagram of the relabeling method \label{fig1}}

\subsubsection{Labeling the slide windows based on the partial data}

The previous labeling method supplies a reasonable method to label the slide windows. However, the larger the window sizes are, the longer the delay will be. In contrast, if we label the slide windows based on the partial sample data in the windows, the problem can partly be solved. Concretely, we use the last three or five sample data in the sliding window to label this sliding window, as shown in Figure \ref{fig2}. In this vein, the delay has been reduced.


\Figure[ht!](topskip=0pt, botskip=0pt, midskip=0pt)[width=3.3in]{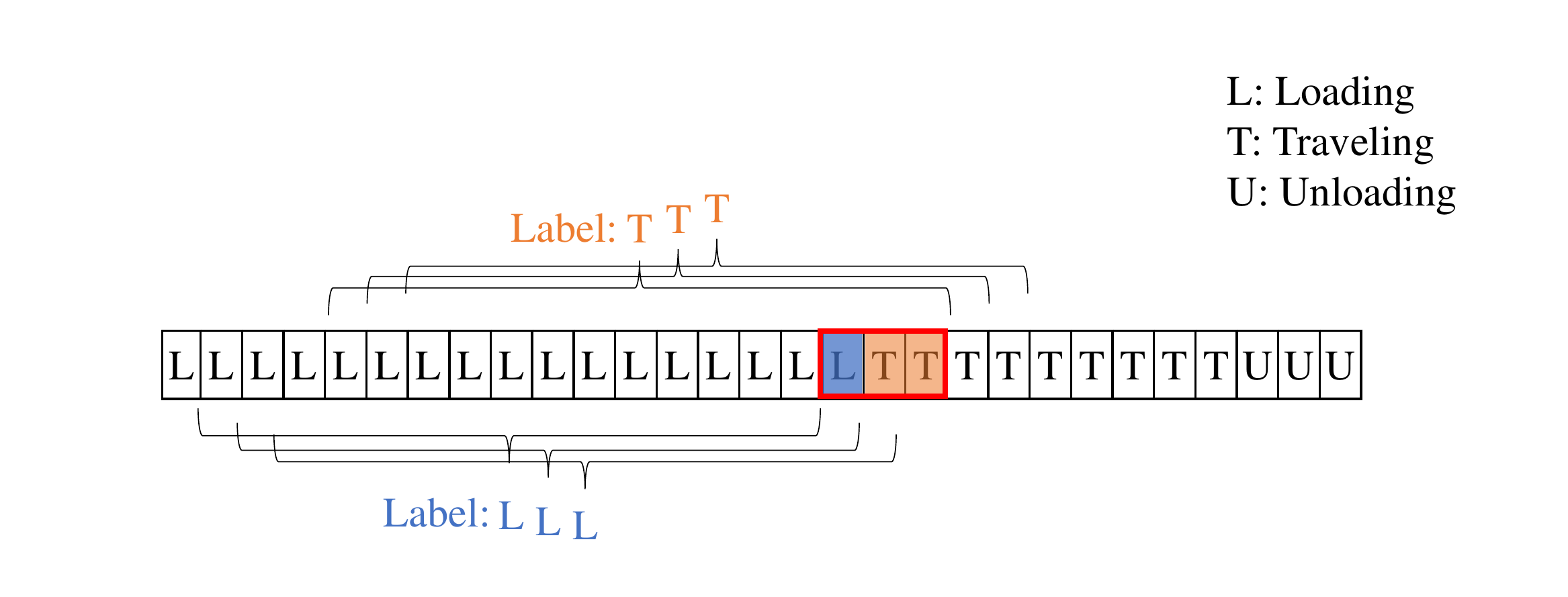}
{The diagram of the relabeling method \label{fig2}}

\section{Comparison bewteen CRDNN and other SOTA time series detection neural networks}

Before we explore the benefits of transfer learning, we should first determine which neural networks should be used as the base network. As mentioned in section II,  LSTM-FCN is considered as a SOTA solution for TSC tasks. In this section, we would like to compare our CRDNN with LSTM-FCN with respect to micro F1, training time, and test time. Here the training time indicates whether the algorithm is suitable for immediately fine-tuning on the working site. The test time shows if the algorithm is appropriate for realtime detection. Our base networks were trained on Nvidia GEFORCE GTX 1050 GPU. In order to find the global minimum rather than the local minimum, we use early stop and set the patient to 100, which means the training process will be stopped 100 epochs after finding the best predictor. To further avoid overfitting, we adopt the L2 regularization method the same as our previous study. The optimizer we used is ADAM \cite{Kingma.2015}. Also, we use ReLU as our activation function since it can be trained faster as Sigmoid. In Tab. \ref{tab: different algorithms}, we demonstrate the performance of different neural networks with different window sizes. Here we use the previous dataset to perform the process of selection of the base networks so that the selected base network can be directly used in the next section where the performance of transfer learning will be discussed. If the model mispredicts the unloading process into the loading process or in reverse, a complicated operation strategy must be designed. Therefore, we only select the models which do not make mistakes in classifying the loading state into the unloading process or in reverse.  Among them, CRDNN with 2 LSTMs with WS 15 has the shortest training time and test time. The training time is 310.19 seconds. Compared to LSTM-FCN with WS 15, it needs only one third training time. Although the micro F1 is slightly worse than LSTM-FCN with WS 15, less than 1\%, we believe than a much shorter training time conducive to a better performance in transfer learning with respect to efficiency.  Moreover, in case that we want to increase the micro F1, we can either increase the WS, or use the other variances of CRDNNs, the one with bidirectional LSTM, to achieve the almost the same micro F1, whose difference is less than 0.1\%. Notice that we do not further pursue to increase the micro F1 since 98\% is already the human-level performance, and thus a further increment might not make sense. Interestingly, although the micro F1 increases as the WS increases, the training time does not always increase as the WS increases.
In short, based on the training results, the LSTM-FCN has a slightly better performance than the CLDNN with 2 LSTM layers and CLDNN with both one bidirectional LSTM layer and one LSTM layer; however, the training time of the LSTM-FCN is enormous pressure for the ECU when we make a transfer learning on the ECU. Thus, we select CRDNN with 2 LSTM layers as our base networks for transfer learning. Also, we select the WS as 15 according to the training results.
 
\begin{table*}[!ht]
	\caption{The performance of the five network structures in respect of total training time (s), Micro F1 (\%), average test duration (ms), and whether it can never mistake an unloading into loading or in reverse}
	\centering
	\begin{tabular}{c|cccc}
	\hline \hline
	 
	 &5
	 &9
	 & 15
	 & 25\\ \hline
     CRDNN (1 LSTM)
     &419.69/ 93.92/ 0.0606/ No
     &214.99/ 96.65/ 0.0703/ No 
     &271.25/ 97.42/ 0.1008/ No
     &248.07/ 97.48/ 0.1338/ No \\
     CRDNN (2 LSTMs)
     &500.00/ 93.65/ 0.0835/ No
     &250.38/ 96.36/ 0.1112/ No
     &310.19/ 97.21/ 0.1457/ Yes
     &410.01/ 98.25/ 0.2014/ Yes \\
     CRDNN (1 LSTM, 1 BiLSTM)
     &370.28/ 92.41/ 0.1111/ No
     &302.89/ 96.13/ 0.1645/ No
     &385.26/ 97.31/ 0.2261/ Yes
     &479.69/ 98.34/ 0.3690/ Yes\\
     CRDNN (2 LSTMs, SAE)
     &768.97/ 95.21/ 0.1350/ No
     &442.94/ 96.88/ 0.1679/ No
     &444.75/ 97.01/ 0.2041/ Yes
     &518.00/ 98.17/ 0.2671/ Yes\\
     LSTM-FCN
	 &1095.00/ 96.17/ 0.1579/ No
     &1270.25/ 98.16/ 0.1698/ No
     &882.13/ 98.14/ 0.1663/ Yes
     &747.39/ 98.32/ 0.2029/ Yes \\
	\hline \hline
	\end{tabular}
	\label{tab: different algorithms}
\end{table*}

\section{Transfer learning based CRDNNs}

Since we do not change the model architecture, there are two potential transfer learning methods: either we can freeze the former parts of CRDNN and only further train the fully connected layers to save the training time, or we can use the pre-trained model's weights as the initial parameters for the further training of the total model. Obviously, the first vein is faster and can mitigate the ECU computational effort. Yet the second way may achieve a better recognition performance. Generally speaking, we can only use the newly gathered data as the validation set, just like other transfer learning tasks did. However, from the users' view, we evaluate the performance both on previous data ($\mathcal D_s$) and new data ($\mathcal D_t$).  To evaluate the accuracy of each approach, we first show the micro F1 value and then illustrate the confusion metrics. Furthermore, to indicate whether a method is suitable for on-site transfer learning, we judge the approaches based on their training time (back-propagation) and test time (forward-propagation). Here we show the training time and test time on a CPU core i7 4720HQ@ 2.6GHz since the results are more appropriate to be used as the benchmark for the onboard ECU. The hyper-parameters and the architecture are shown in Tab. \ref{tab_2}, and the results are shown in Tab. \ref{tab: transfer learning}, where the ND, PD, FS, FTF, OTF denotes newly gathered dataset, previous dataset, training from scratch, fully connected layers transfer learning, and overall transfer learning. For transfer learning, we reduce the patient to 50 for the purpose of achieving a relatively faster training process. 

\begin{table}[!ht]
	\centering
	\caption{Parameters of CRDNN with 2 LSTMs}
	\begin{tabular}{l|cl}
	\hline \hline
	Hyper-parameters
		& Value \\ \hline
	Hyper-parameters
		& 15 \\
	Batch size
		& 128 \\
   	Initial learning rate (decay during learning)
		& 1x$10^{-4}$ \\
	Num filter conv1D
		& 10 \\
	Kernel size 
	    & 5 \\
	Num units 1$^{st}$ layer (RNN) 
	    & 32 \\
    Num units 2$^{nd}$ layer (RNN) 
        & 32 \\
    Num units 1$^{st}$ layer (DNN) 
        & 32 \\
    Num units 2$^{nd}$ layer (DNN)
        & 32 \\
	\hline \hline 
	\end{tabular}
	\label{tab_2}
\end{table}

\subsection{Training from the scratch as benchmark (ND+PD+FS)}

In order to have a basic overview between the CRDNN trained from the scratch and the CRDNN trained by means of transfer learning, we demonstrate the training process of the  CRDNN trained from the scratch and the CRDNN with the method transfer learning, separately. The CRDNN from scratch is used as the benchmark to illustrate the benefits of transfer learning. By means of training from scratch, each epoch contains 143 Y cycles data since we mixed the newly gathered and the previous dataset together, and the training process can stop at about 75 epochs, as shown in Fig. \ref{fig:fig8} (a). 

\subsection{Only further train the FCN (ND+FTF)}

\begin{figure*}[!t]
        \newcommand{\w}{0.4}
        \centering 
        \subfloat[][CM train from the scratch with previous and new data]{
            \includegraphics[width=\w\textwidth]{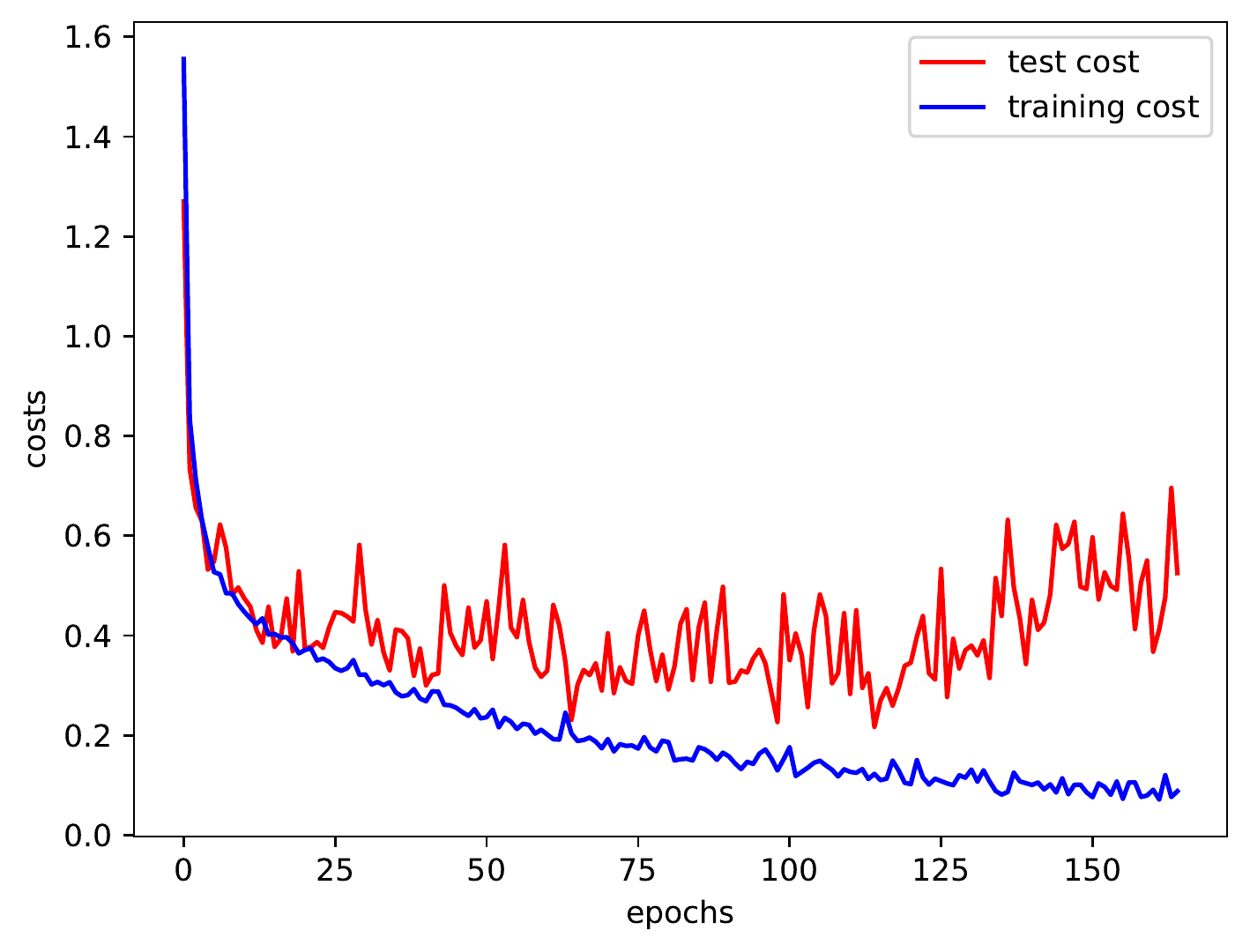}}
            \hfil 
        \subfloat[][CM trained from scratch with new data]{
            \includegraphics[width=\w\textwidth]{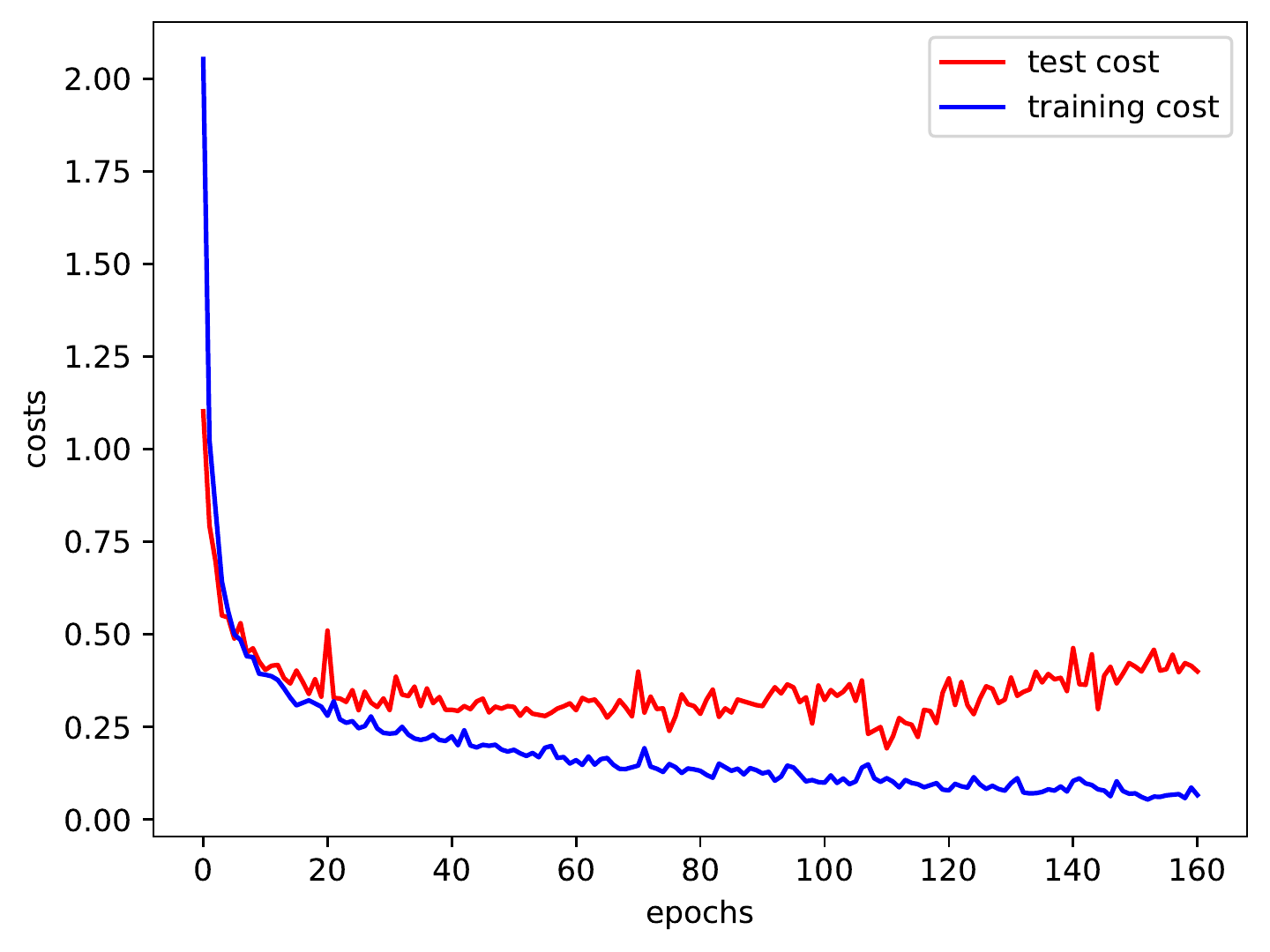}}
            \hfil 
        \subfloat[][ \centering CM trained with method transfer learning: only the final fully connected layers are trainable]{
            \includegraphics[width=\w\textwidth]{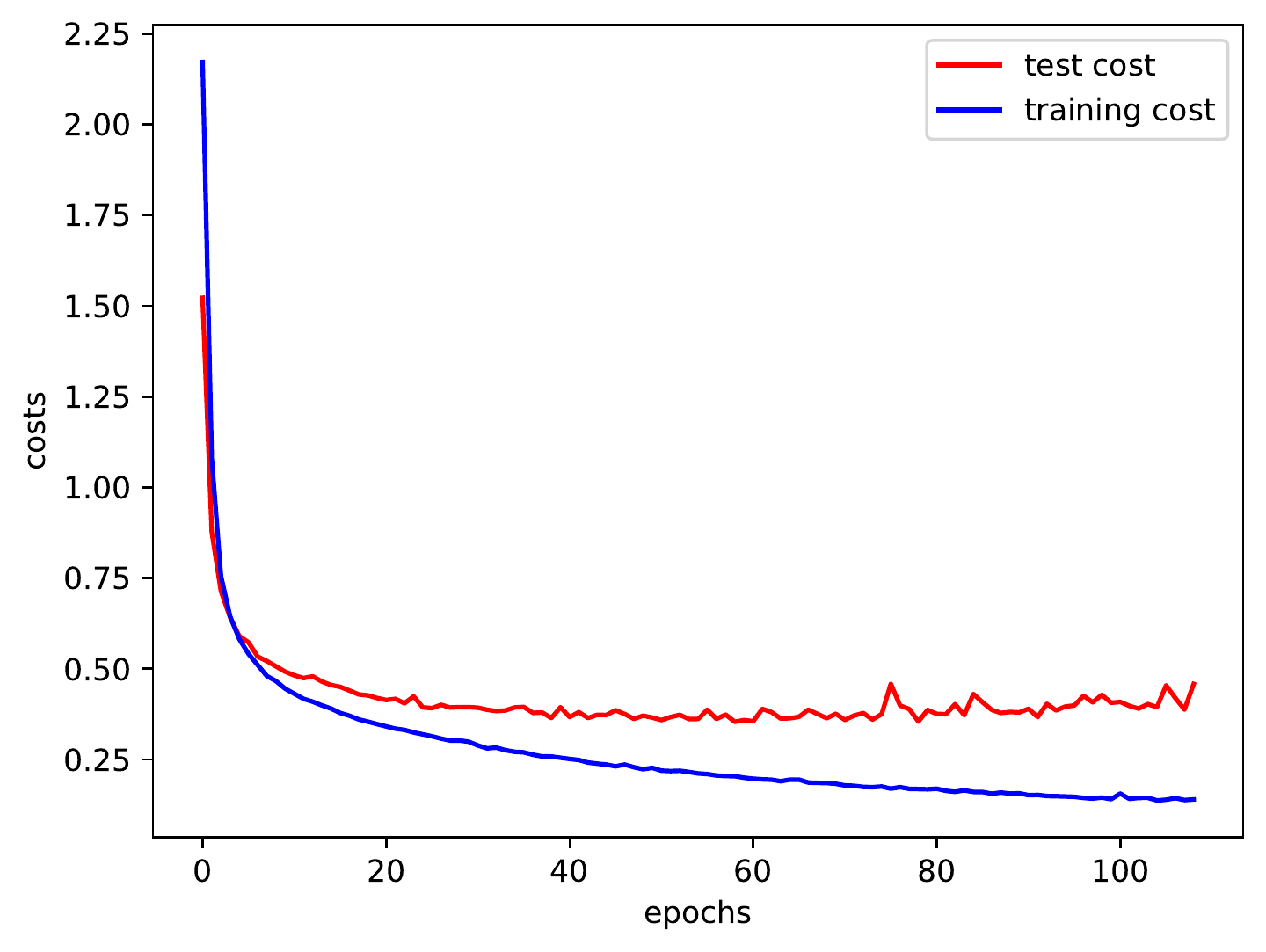}} 
            \hfil 
        \subfloat[][\centering CM trained with method transfer learning: all layers are trainable]{
            \includegraphics[width=\w\textwidth]{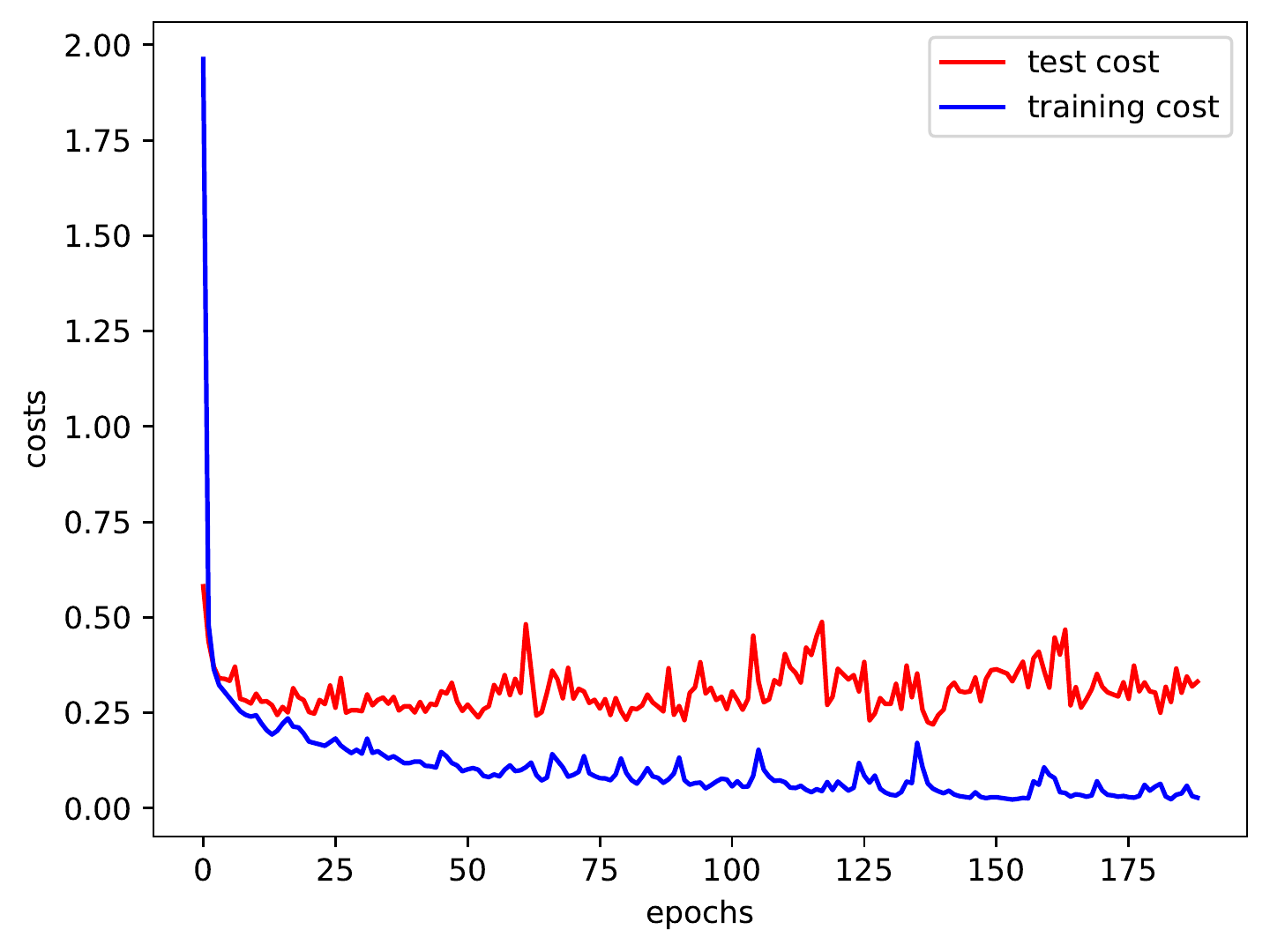}} 
        \caption{Cost versus epoches of CRDNNs from scratch and with transfer learning} \label{fig:fig8}
\end{figure*}

\begin{table*}[!ht]
	\caption{The performance comparison between different training methods}
	\centering
	\begin{tabular}{c|ccccc}
	\hline \hline
	
	 &F1 on ND
	 &F1 on PD
	 & Sample per epoch 
	 & Trainable parameters
	 & Training time (s) \\ \hline
	 ND + FS
	 &0.9730
     &0.8432
     &  5695
	 & 16295
	 & 174.70\\
     ND + FTF
     &0.9540
     &0.8305
     & 5695 
	 & 2211
	 & 62.90\\
     ND + OTF
     &0.9798
     &0.9038
     &5695
     &16295
     &224.04\\
     ND + PD + FS
     & 0.9655
     & 0.9664
     &32473
     &16295
     &668.30\\
	\hline \hline
	\end{tabular}
	\label{tab: transfer learning}
\end{table*}

\begin{figure*}[!t]
        \newcommand{\w}{0.4}
        \centering 
        \subfloat[][CM train from the scratch with previous and new data]{
            \includegraphics[width=\w\textwidth]{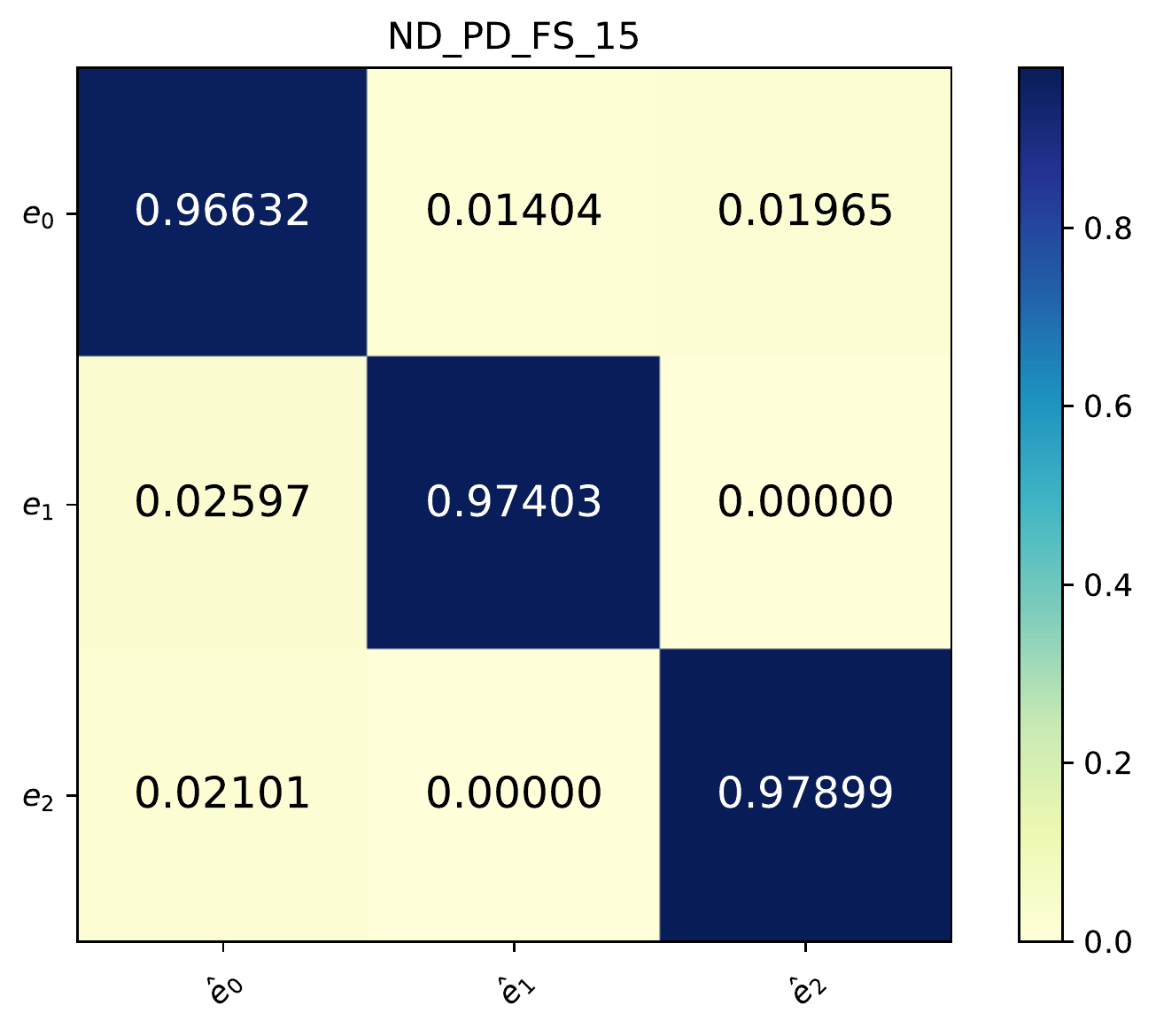}}
            \hfil 
        \subfloat[][CM trained from scratch with new data]{
            \includegraphics[width=\w\textwidth]{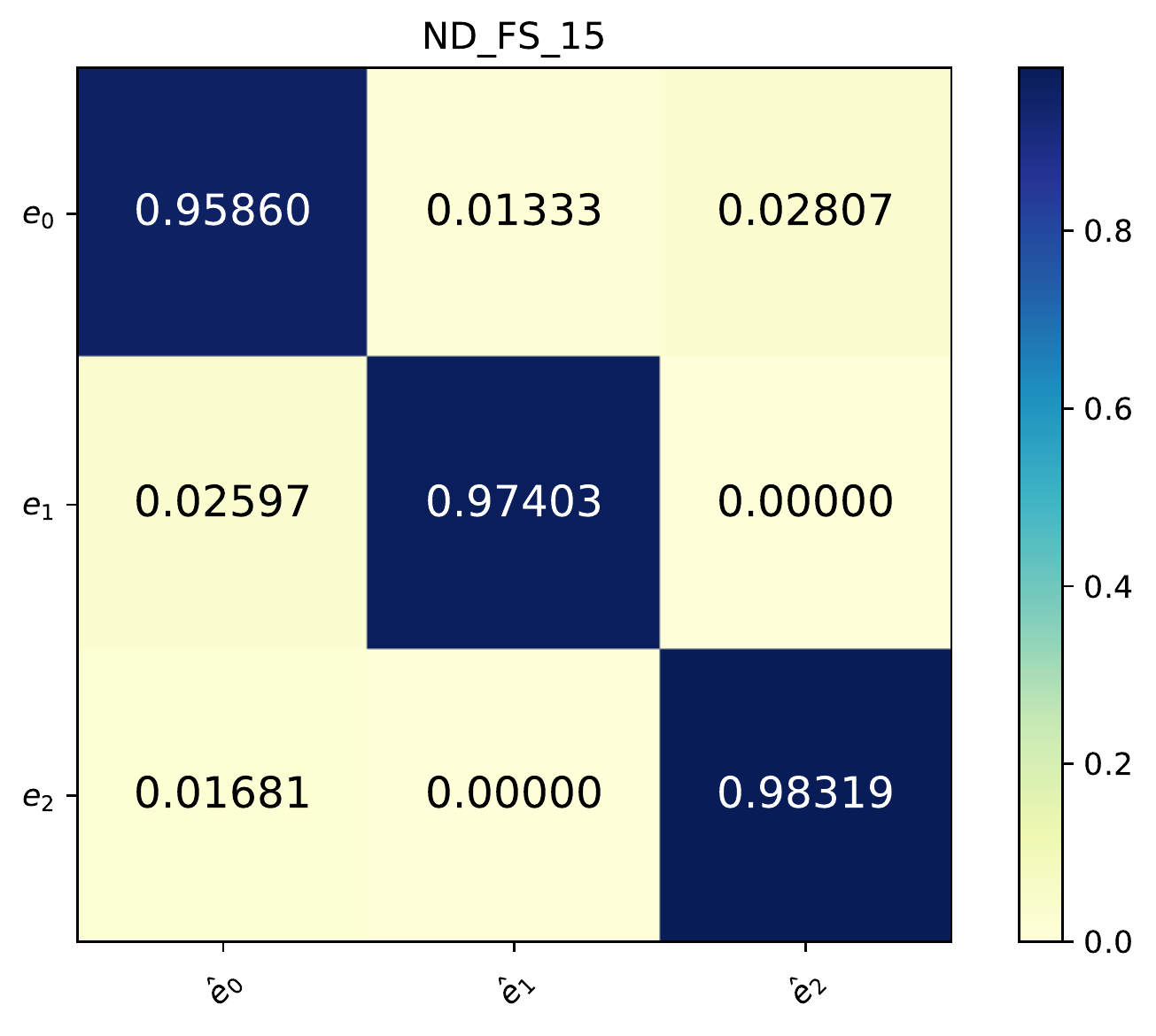}} 
            \hfil
        \subfloat[][ \centering CM trained with method transfer learning: only the final fully connected layers are trainable]{
            \includegraphics[width=\w\textwidth]{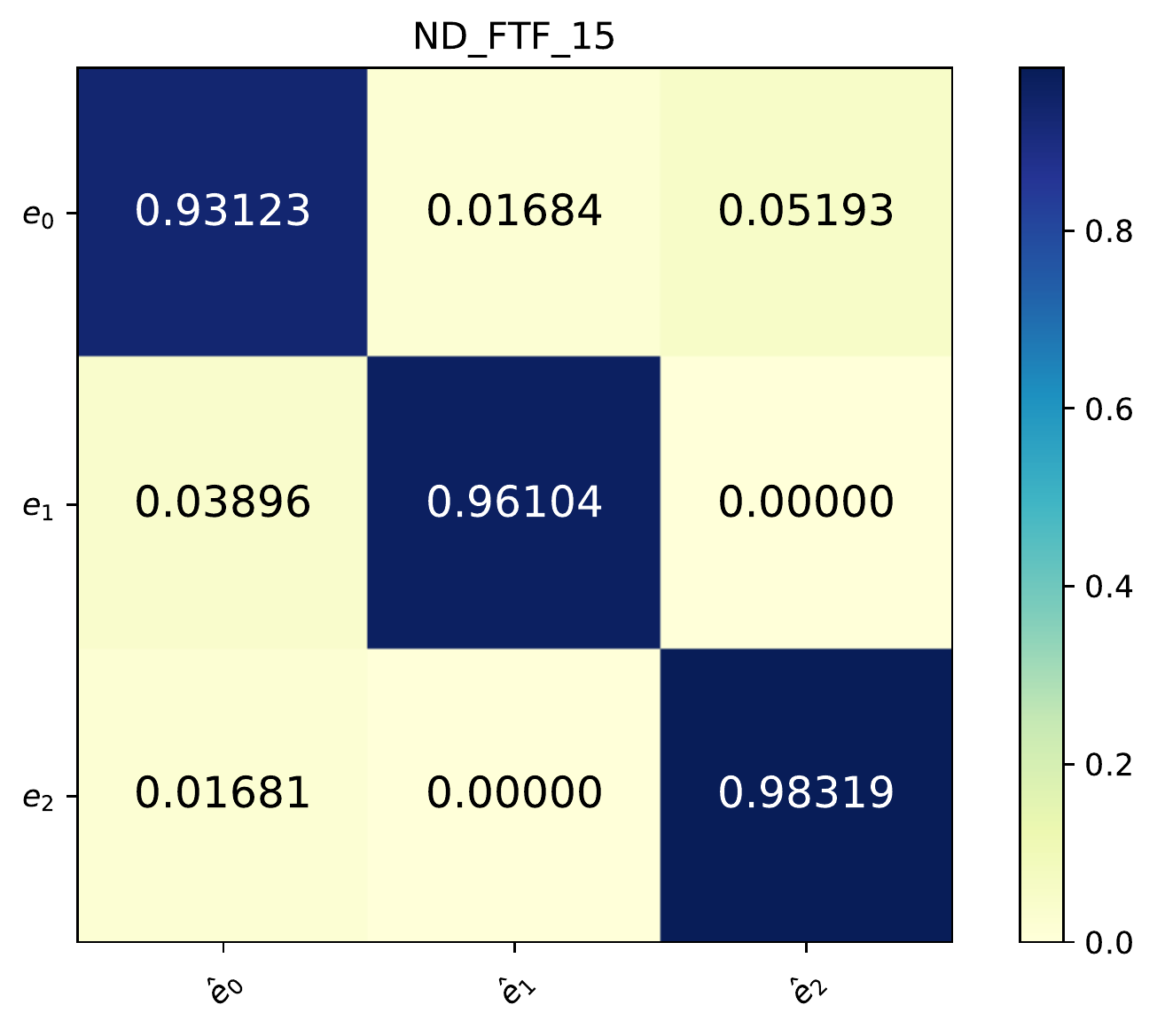}} 
            \hfil
        \subfloat[][ \centering CM trained with method transfer learning: all layers are trainable]{
            \includegraphics[width=\w\textwidth]{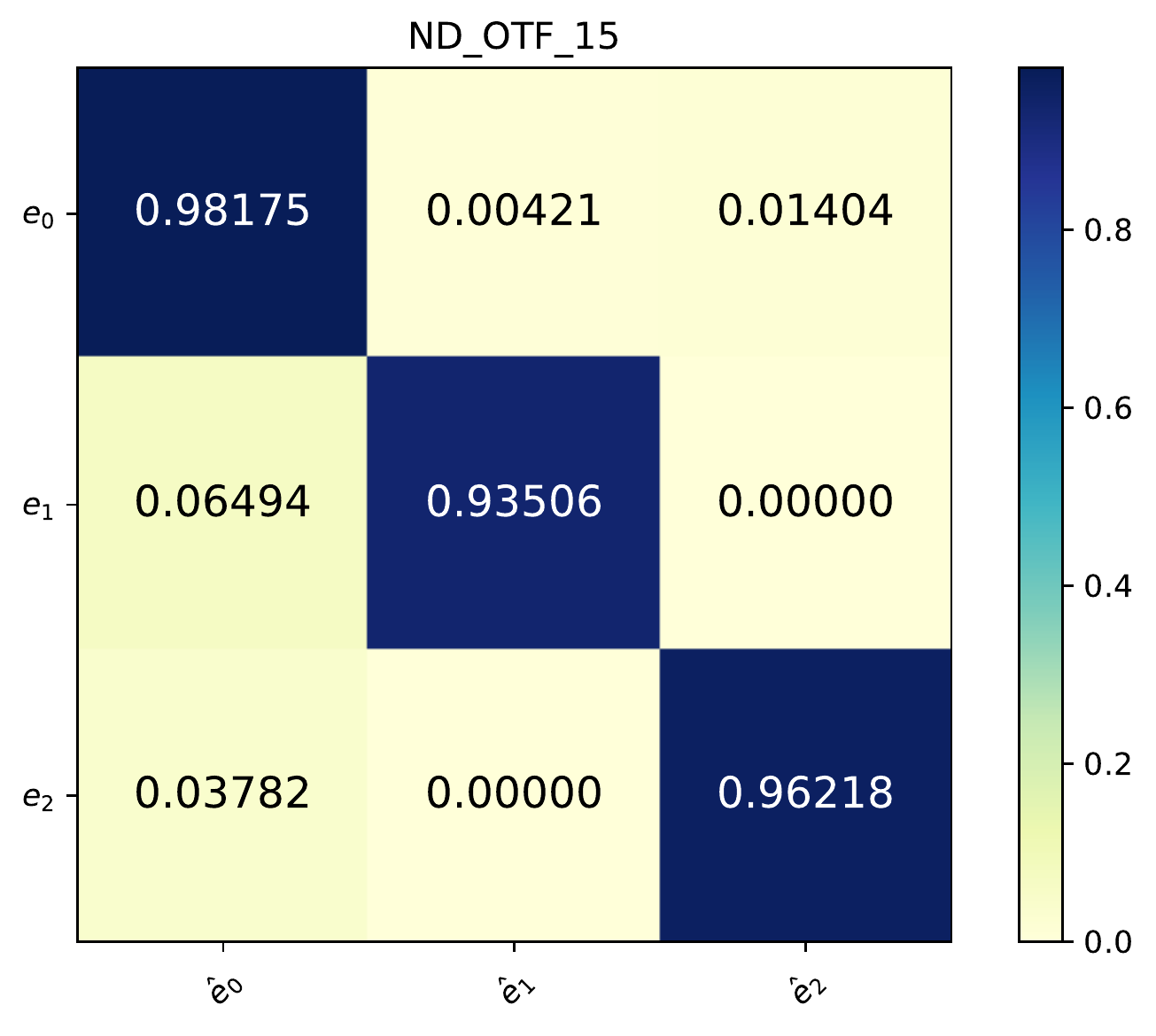}} 
        \caption{ Confusion matrices of CRDNNs from scratch and with transfer learning} \label{fig:fig9}
\end{figure*}


Because we can train the DNNs faster than CNNs, we firstly only train the final fully connected layers in the CRDNN, and then analyze the performance. 
As we can see, the model is further trained with the new dataset. Here each epoch has only 24 Y cycles, and the training process stops at about 60 steps. After the transfer learning process, we can see that the prediction accuracy is much better than the results shown in Fig. \ref{fig: problem}.  
Concretely, the micro F1 is increasing to about 95\%. However, we can utter that the results are satisfying but not perfect as the totally retrained CRDNN. We are observing Fig. \ref{fig:fig8} (c), the current neural network is lack of learning ability for further improving the performance of the CRDNN since the validation cost does not change during the training cost goes down.



\subsection{Train the total part of CRDNN (ND + OTF)}

Fig. \ref{fig:fig8} (d)  is the result when we further train all the parts of CRDNN with the newly gathered data. Obviously, the CRDNN has a stronger learning capability compared to CRDNN with FTF since the test cost goes down deeper as the epoch increases. The micro F1 of the CRDNN with OTF is higher since the state traveling occupies a majority of our dataset. In order to let the newly trained model can also have a good performance on the previous data, we introduced the soft weight sharing method that uses different learning rates for different layers of neural networks. Concretely, we let the learning rate for the CNNs and RNNs smaller as the DNNs. 


\subsection{Evaluation the benefits of transfer learning}

The performance of these four methods is shown in Tab. \ref{tab: transfer learning}. Since the first three methods are trained on the newly gathered data, the samples per epoch are much fewer than then fourth methods. Also, in case that we only train the fully connected layers, the trainable parameters are the fewest. Both of them are good for reducing training time. Thus, the training time for ND+ FTF can reduce to one-tenth (10\%) compared to ND+PD+FS, and one third (35\%) compared to ND+FS. Here the training time is 62.90s; However, since the model can be trained on different onboard ECU or smartphone, the concrete numbers shown in our table are only made sense to be used as a benchmark to compare the performance of one approach to the other approaches. For instance, some ECUs on the mobile machines may have a relatively lower computational capability resulting in 10 times longer training time than the value here shown. Also, it is possible that the onboard ECU is even faster than this training time because mobile machines usually have a powerful energy source. Based on the comparison, we use the method ND+FTF as an emergency method to let the model can work with high accuracy immediately on the new task after new labeled data are fed into the model. ND+FS shows a great accuracy on the new data; however, since the newly gathered dataset is relatively small, the generalization capability of this approach is suspicious. The other transfer learning method, overall transfer learning, has the best performance on new data. It is also good at detecting the previous data, which indicates that it has a good generalization capability.
Moreover, the training time is only one third (33\%) compared to the ND+PD+FS. Therefore, we recommend using ND+FTF to train the network in the case that it is not so hurried or the mobile machine has a relatively powerful ECU.  Note that the micro F1 of ND+PD+FS is slightly worse than the results in  Tab. \ref{tab: different algorithms} because the patient is fewer. As shown in the confusion matrices in Fig. \ref{fig:fig9},  the models do not mistake the loading process with the unloading process, which denotes that all the models can mitigate the design of operation strategy; thus, all of them have the potential to be used, in case that OEMs have their special wish.


\Figure[ht!](topskip=0pt, botskip=0pt, midskip=0pt)[width=2.9in]{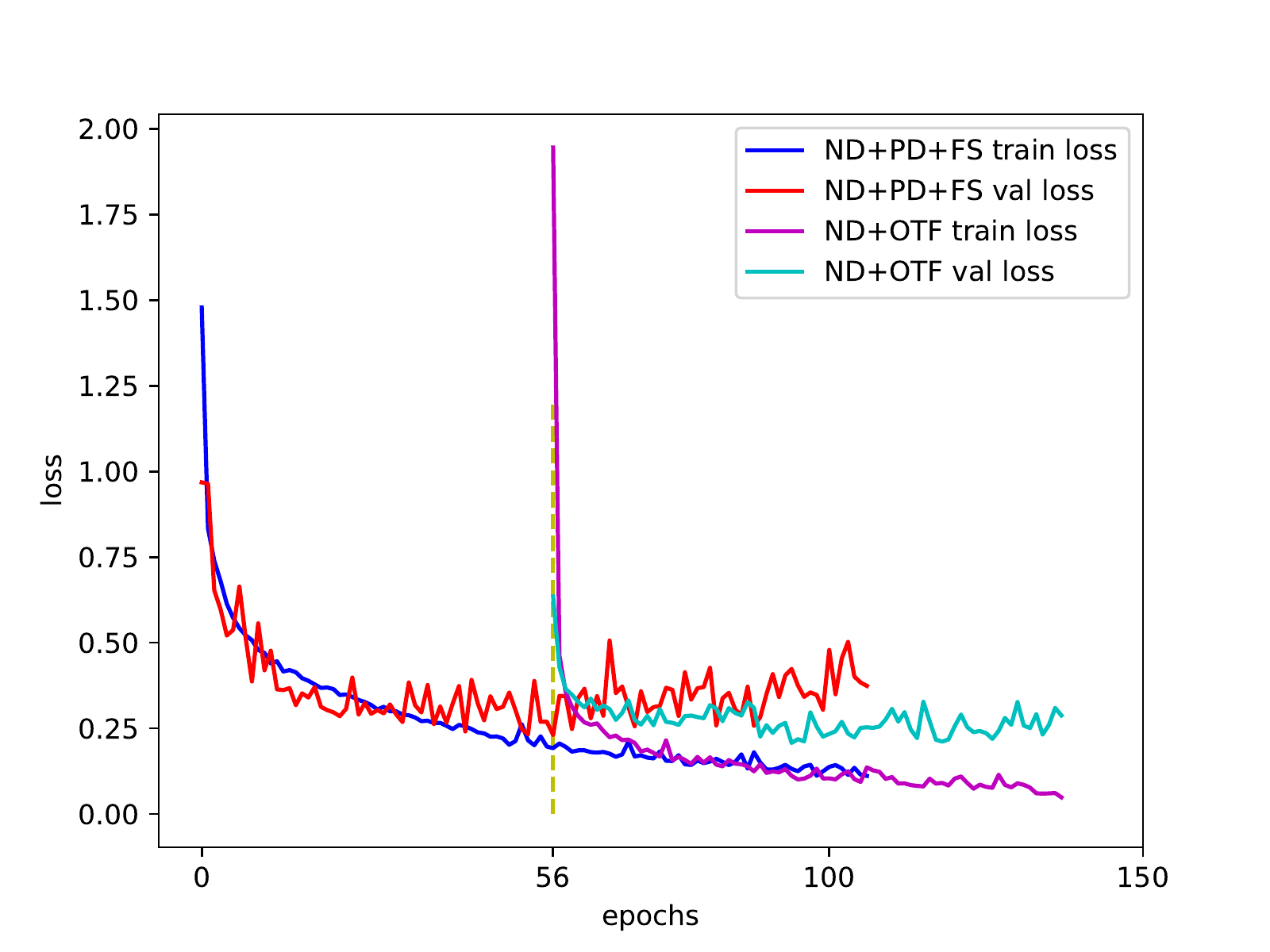}
{The mechanism of transfer learning. The blue line is the training cost on ($\mathcal D_s$), the red line is the validation cost ($\mathcal T_s$), the purple line is the training cost on ($\mathcal D_t$), and the cyan line is the validation cost ($\mathcal T_t$) \label{fig: cost evaluation}}

To illustrate the mechanism of time saved due to offline learning with online adaption compared to pure online learning, we show the training process of transfer learning. As shown in Fig. \ref{fig: cost evaluation}, the training and validation cost on precious explode at the epoch 63. The base network on the computer is finished at step 63 since the validation cost begins to grow. Hence, at this time point, we added the new labeled dataset to simulate the real scenario for transfer learning. Right after the new data are considered, both training and validation cost goes to an extremely high level since the dataset has an enormous variance.
Consequently, the prediction results must be unsatisfying. Interestingly, only after a few steps of further training on the on-board ECU, the cost goes dramatically down to the low level. As a result of that, the CRDNN is again suitable to predict the truck loading process even when the scenario is quite different from the original dataset.

Based on the results of this section, we can say that transfer learning is a powerful tool to let the CRDNN be robust to the challenging Y cycles detection tasks. The transfer-learning based CRDNN with 2 LSTMs is the most appropriate model since it can be retrained much faster than LSTM-FCN with only 1\% accuracy lost. Without transfer learning, the model can not guarantee excellent performance for the new target task ($\mathcal T_t$); thus, we recommend using transfer-learning-based CRDNN for the detection of Y cycles.

\section{The advantages of this system from engineers' view}
Here we would like to sum up the main advantages of the transfer-learning based CRDNN and the corresponding IoT system as strong, fast, and easy.

\subsection{Strong}
This system is aimed to improve the efficiency of the novel torque-controlled hydrostatic mobile machines by correctly detecting the working process. This system can automatically recognize the working state without an additional button or human action, which offers essential information for the energy regeneration process. Thanks to the transfer learning, the system can be adapted to a new machine, even where there has a different distribution as the source dataset, without a complicated calibration process. The test accuracy of this working state recognition system can reach 98\% on the challenging dataset \cite{Xiang.2020d}, which achieves the human-level performance and guarantees accurate recognition.
The strong ability of generalization of transfer-learning-based CRDNN is proven.

\subsection{Fast}
Usually, an excellent ability of generalization is based on the sacrifice of speed. However, the transfer-learning-based CRDNN is fast. It is an offline method with online adaption; thus, it is a realtime algorithm. Also, transfer learning needs much less computation effort resulting in the on-site training capability of CRDNN. 

\subsection{Easy}
Generally speaking, an interface that controls an extensive system is complicated. However, the IoT system designed in this paper is easy. It is an APP on the iPhone. The operators only need to give the data the appropriate label and check the model accuracy. The system automatically does most of the training steps. 

At the end of this section, we demonstrate the performance of different approaches in Tab. \ref{tab: performance of different approach}, where the online learning approach was evaluated with the batch size is equal to 1. 

\begin{table}[!ht]
	\caption{Performance comparison of different learning approach}
	\centering
	\begin{tabular}{c|ccc}
	\hline \hline
	 
	 & Offline
	 & Online with adaption
	 & Online\\ \hline
     Real time 
     & +
     &+ 
     &-\\
     Ability of generalization
     &-
     &+
     &+ \\
     Learning ability
     &++
     &++
     &+\\
	\hline \hline
	\end{tabular}
	\label{tab: performance of different approach}
\end{table}

\section{Conclusion}

In this paper, we update the naive CRDNN to the transfer-learning-based CRDNN. Thanks to the transfer learning, the generalization ability of CRDNN has been much enhanced so that it becomes a powerful solution for solving the high variance problem in detecting the truck loading process. Since transfer learning needs new data, we complete the IoT system of mobile machines by building a human-machine communication system on the smartphone for the purpose of gaining the data quickly. The model we recommend can be trained very fast so that the workers can adapt the model directly on the working site after gaining the new data rather than sending the data to the deep learning specialist. As the results showed, we can say that the proposed methods can always help the pre-trained CRDNN to achieve satisfactory performance with respect to precision and recall. Besides, the training time on the onboard ECU can reduce at least about 70\% to 90\% compared to if we retrain the neural network from scratch on the onboard ECU. Also, we use the new method to label the sliding window, so that we can partly solve the delay of the prediction results in the previous version of CRDNN.

\subsection{Outlook}

The purpose of detecting the Y cycles is to improve the holistic efficiency of torque controlled mobile machines. Transfer learning based CRDNN has already proved to solve the most critical pain points in this task. Although some more powerful algorithms may be proposed in the future, they might not bring much more benefits regarding this task. However, the IoT system designed for human-machine communication shall be further developed due to its potential. 

The connected mobile machine is undoubtedly a research focus shortly. While the Bluetooth technology is considered as a cheap and reliable communication solution for human and machines interaction, we believe the next generation communication tools should have access to cellular networks (4G or 5G) since the other components, such as hydraulic pump and hydraulic motor, of the mobile machines also have the requisite to connect to the communication networks for components monitoring, which might overload the Bluetooth. Moreover, we believe the fleet management can facilitate the industry of mobile machines.  Therefore, in the next generation of the connection system, we will take advantage of 5G to achieve a fully connected working site. Thanks to the cloud, CRDNN can be further trained with newly gathered data whenever a customer label the new dataset for their newly developed mobile construction machines and thus become even more reliable.

\EOD

\bibliography{Literature.bib}{}
\bibliographystyle{IEEEtran}

\begin{IEEEbiography}[{\includegraphics[width=1in,height=1.25in,clip,keepaspectratio]{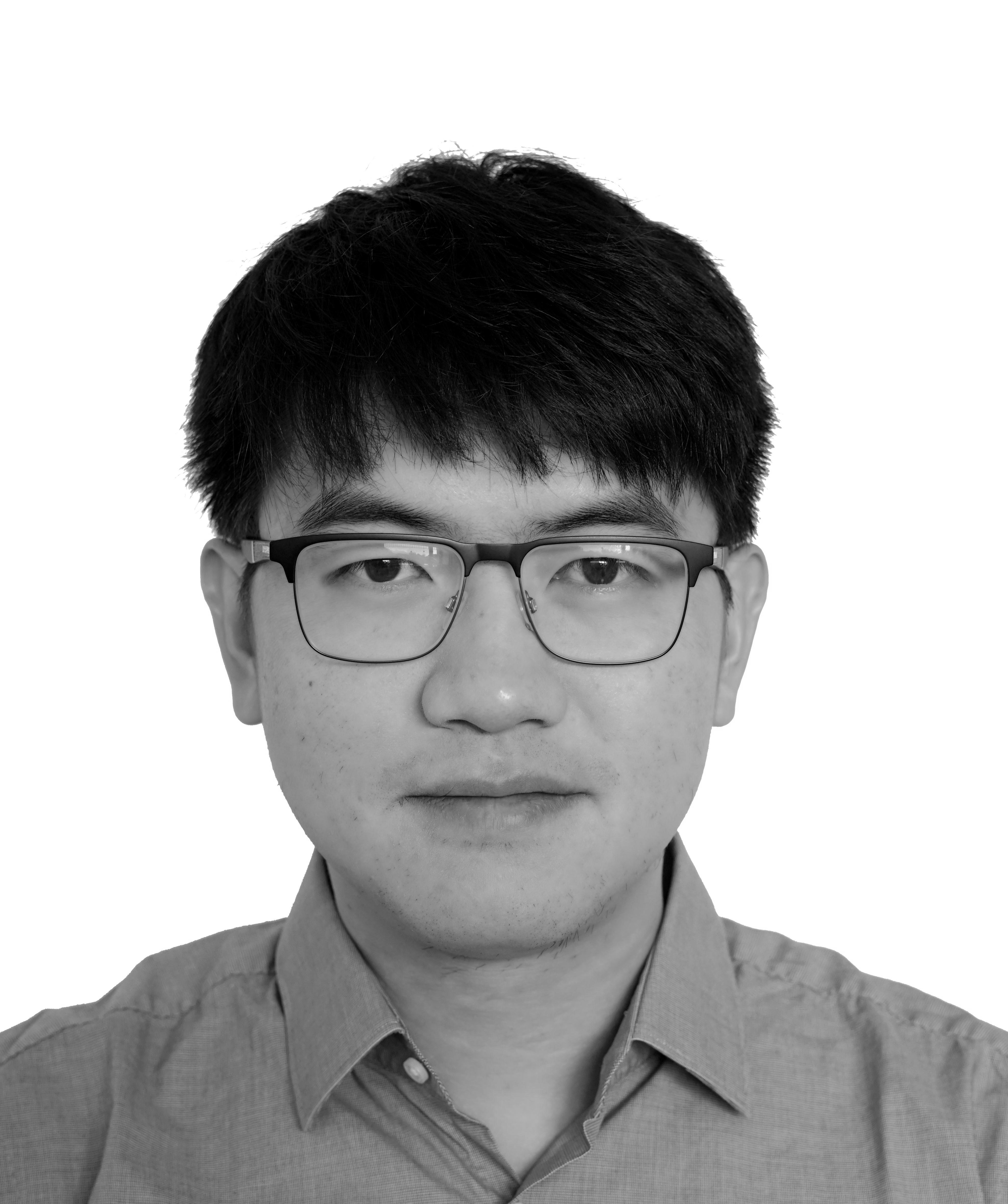}}]{Yusheng Xiang} is pursuing his PhD degree at Institute of Vehicle System Technology, Karlsruhe Institute of Technology, Karlsruhe, Germany. Also, he is a research scientist at Robert Bosch GmbH, Germany. From Sep. 2020, he is a visiting scholar at University of California, Berkeley, USA. He received M.Sc. degree in Vehicle Engineering with the focus of Mathematical Model Building and Simulation from the Karlsruhe Institute of Technology, Karlsruhe, in 2017. He has authored 6 influential journal and international conference papers, and holds 4 patents. His group deals with the improvement of mobile machines performance using Artificial Intelligence and Internet of Things.

\end{IEEEbiography}

\begin{IEEEbiography}[{\includegraphics[width=1in,height=1.25in,clip,keepaspectratio]{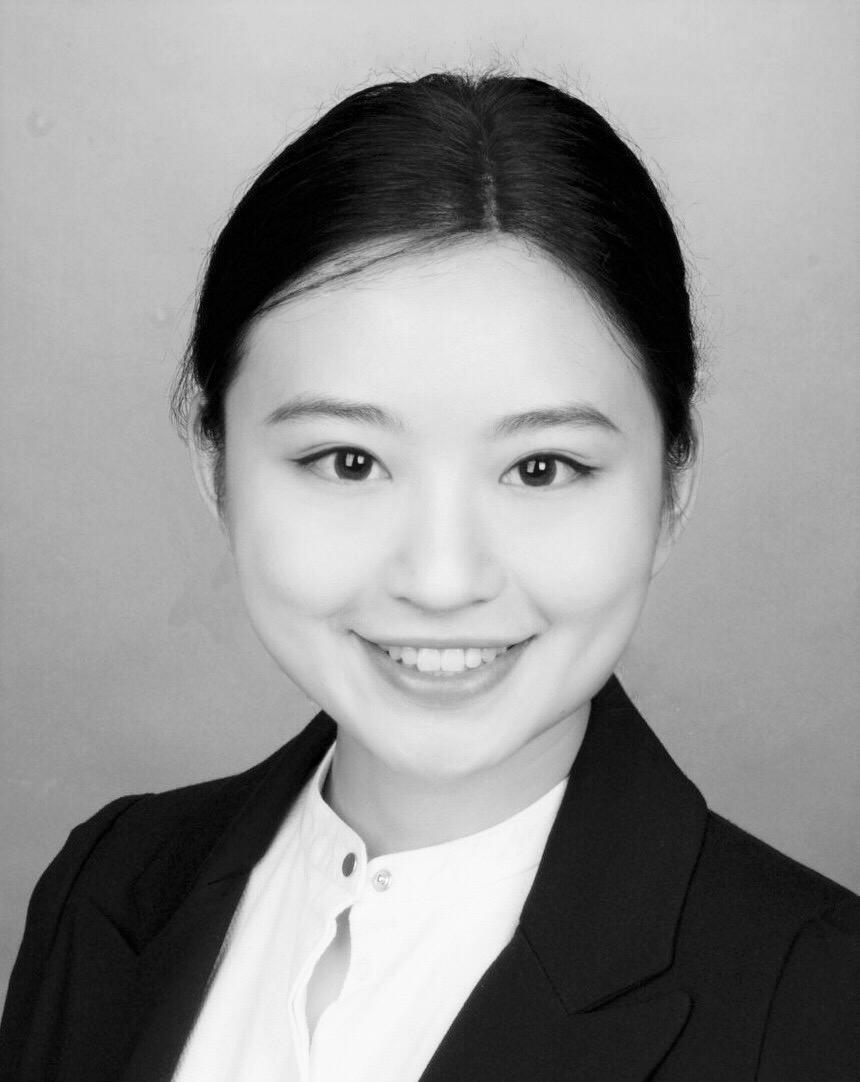}}]{Tianqing Su} received the B.Sc. degree in telecommunication from the Xidian University, Xi'an, China, in 2013, and the M.Sc. degree in Information Systems Engineering from the Technische Universität Braunschweig, Germany, in 2017. She is currently a Software Engineer in the field of hybrid and electric vehicle at Continental AG, Regensburg, Germany. The IoT framework used in this research is based on the research she did during her internship at Robert Bosch GmbH, Abstatt, Germany.

\end{IEEEbiography}

\begin{IEEEbiography}[{\includegraphics[width=1in,height=1.25in,clip,keepaspectratio]{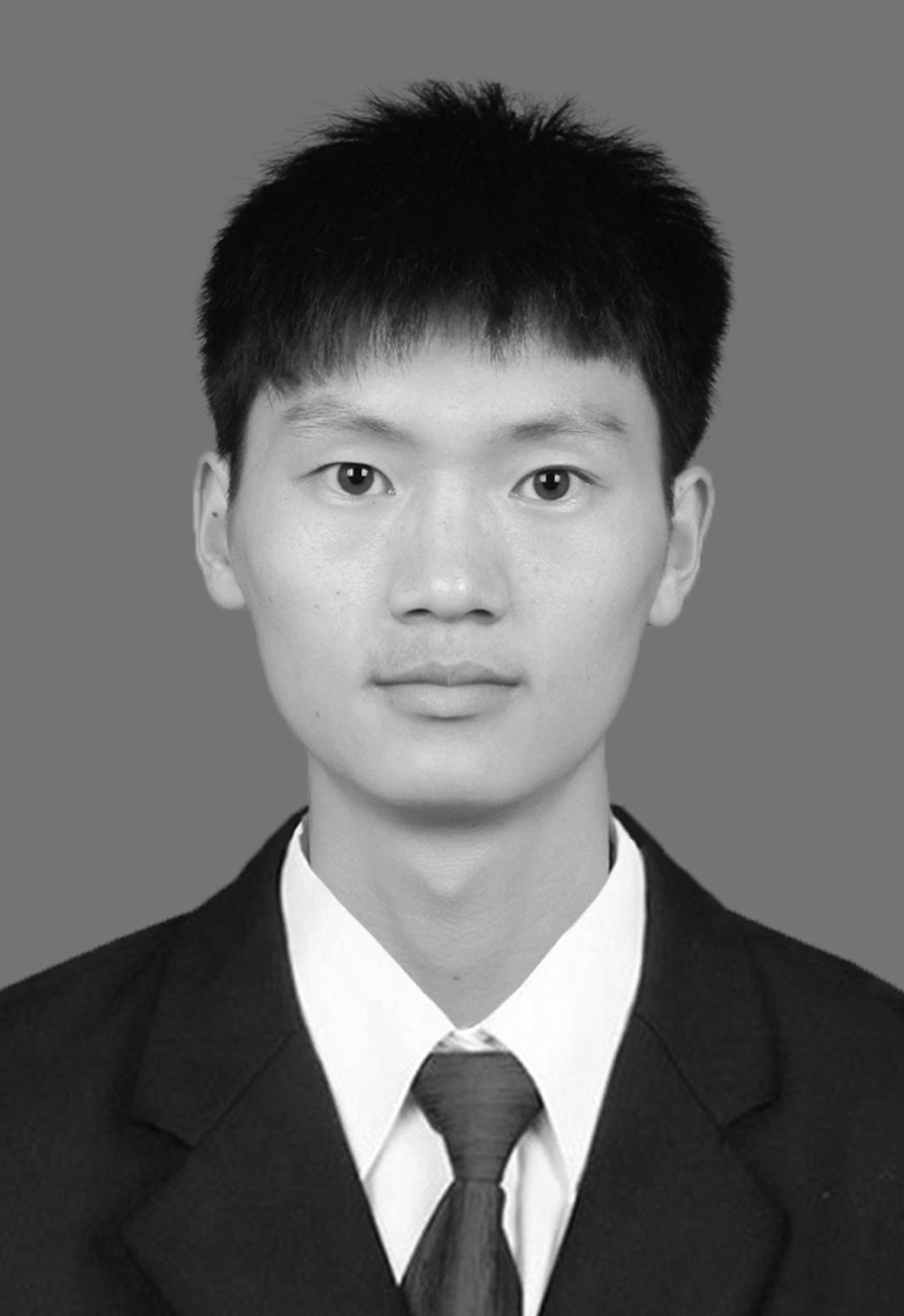}}]{Tian Tang} received M.Sc. degree in Mechanical Engineering with the focus of Deep Learning from the Karlsruhe Institute of Technology, Karlsruhe, Germany, in 2020. He has also received M.Sc. degree in Mechatronics Engineering with the focus of fluid simulation from the Southwest Jiaotong University, Chengdu, in 2019.

\end{IEEEbiography}

\begin{IEEEbiography}[{\includegraphics[width=1in,height=1.25in,clip,keepaspectratio]{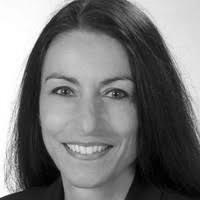}}]{Christine Brach} is the department leader of System Engineering in the division of mobile hydraulics at Robert Bosch GmbH where she leads interdisciplinary teams for large system development projects. As a supervisor for 3 PhD students, she has 15 publications in journals, conferences and patents. 

\end{IEEEbiography}

\begin{IEEEbiography}[{\includegraphics[width=1in,height=1.25in,clip,keepaspectratio]{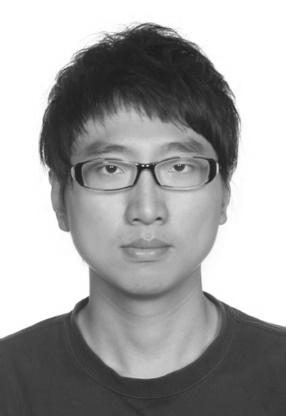}}]{LiBo Liu} received the B.Sc. degree in electrical engineering from Shanghai Jiao Tong University (SJTU) in 2012 and the dual master’s degree in the same major from SJTU and Berlin Institute of Technology (TUB) in 2015. Since 2017, he has been working toward the Ph.D. degree under the scheme of eMobility doctoral program between Robert Bosch GmbH and Ulm University in Germany.
His research interests include the cascaded multilevel converter/inverter for electric vehicle (EV) applications, low-harmonic low-loss modulation scheme for high-speed electrical drive, and stationary battery storage system.

\end{IEEEbiography}

\begin{IEEEbiography}[{\includegraphics[width=1in,height=1.25in,clip,keepaspectratio]{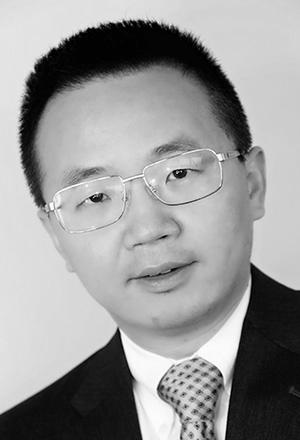}}]{Samuel Mao} received his Ph.D. degree from the University of California at Berkeley in 2000. After that, Prof. Mao started his career at Lawrence Berkeley National Laboratory, where he was a career staff scientist until 2013. He returned to U.C. Berkeley campus as an adjunct professor in 2004, when he also established the Clean Energy Engineering Laboratory that has spun off an international technology development and commercialization institution, the Institute of New Energy, launched in 2013. Having published 180 refereed research articles that have received more than 43,000 citations, Prof. Mao is also an inventor of 80 patents in the U.S. and abroad. He delivered nearly 100 invited, keynote or plenary speeches at international conferences, and previously served as a technical committee member, program review panelist, grant proposal evaluator, and national laboratory observer for the U.S. Department of Energy. In addition to co-founding three international materials and energy technology conferences, he co-chaired Materials Research Society (MRS) annual meeting in the spring of 2011, and the International Conference on Clean Energy in 2012. Prof. Mao received an “R\&D 100” Technology Award (2011) for his technological innovation, and a Berkeley MEGSCO Faculty Teaching Award (2008) for his dedication to higher education.

\end{IEEEbiography}

\begin{IEEEbiography}[{\includegraphics[width=1in,height=1.25in,clip,keepaspectratio]{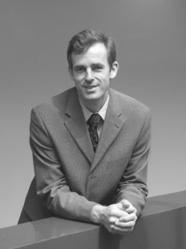}}]{MARCUS GEIMER} received his Diploma degree in mechanical engineering from the RWTH Aachen University, Aachen, Germany in 1990. In 1995, he received his PhD from the Institute of Hydraulics and Pneumatics, today named Institute for Fluid Power Drives and Systems, RWTH Aachen University. He started his industrial career 1995 in the field of construction, where he was the leader of the research group for hydraulic breakers. In 2000, he changed to the hydraulic industry, where he leads the construction and customer development for mobile hydraulics. 
Since 2005, he is a full professor and director at the Institute of Mobile Machines (Mobima), at the Karlsruhe Institute of Technology KIT, Germany. His research activities focus on drives and controls for mobile working machines, like agriculture, construction and municipal vehicles. Research projects on hydrostatic, electric and hybrid drives, as well on traction as on function drives, have been successfully completed. Modern control strategies, like machine learning methods, neural networks or predictive control, are under research.

\end{IEEEbiography}

\end{document}